\documentclass[journal]{IEEEtran}
\usepackage{ifpdf}

\usepackage{amsmath,amssymb,amsfonts}
\usepackage{algorithmic}
\usepackage{graphicx}
\usepackage{textcomp}
\usepackage{wrapfig}
\usepackage{multirow}
\usepackage{color}
\usepackage{booktabs}

\usepackage{cite}

\ifCLASSINFOpdf
\else
\fi
\usepackage{graphicx}

\usepackage{amsmath}
\usepackage{algorithmic}

\usepackage{array}

\usepackage[caption=false,font=normalsize,labelfont=sf,textfont=sf]{subfig}
\usepackage{fixltx2e}

\usepackage{stfloats}
\usepackage{url}

\begin{document}
%
\title{Event-enhanced Passive Non-line-of-sight imaging for moving objects with Physical embedding}
%
%
%

\author{Conghe Wang, Xia Wang, Yujie Fang, Changda Yan, Xin Zhang and Yifan Zuo 
\thanks{This work has been accepted for publication in IEEE. Copyright may be transferred without notice, after which this version may no longer be accessible.}
\thanks{This work was supported in part by the National Natural Science Foundation of China under Grant 62031018. (Corresponding author: Xia Wang)}
\thanks{Conghe Wang was with the Key Laboratory of Photoelectronic Imaging Technology and System of Ministry of Education of China, School of Optics and Photonics, Beijing Institute of Technology, and he is currently with the Department of Electronic Engineering, Tsinghua University.(e-mail: bit1120170840@163.com)}
\thanks{Xia Wang, Yujie Fang, Changda Yan, Xin Zhang and Yifan Zuo are with the Key Laboratory of Photoelectronic Imaging Technology and System of Ministry of Education of China, School of Optics and Photonics, Beijing Institute of Technology. (e-mail: angelniuniu@bit.edu.cn, ycd$\_$950601@163.com, zx1027533025@163.com, zuoyifan$\_$bit@outlook.com , fyj0202@126.com).} 
}

%
%

\markboth{Journal of \LaTeX\ Class Files, June~2024}%
{Wang \MakeLowercase{\textit{et al.}}: Event-enhanced Passive Non-line-of-sight imaging for moving objects with Physical embedding}
%



\maketitle

\begin{abstract}
Non-line-of-sight (NLOS) imaging with intelligent sensors emerges as a novel technique in imaging and sensing occluded objects around corners. With the innovation of bio-inspired neuromorphic sensors, the applications of novel sensors in unconventional imaging tasks like NLOS imaging have shown promising prospects in intelligent perception, encompassing autonomous driving, medical endoscopy and other sensing scenarios. However, the most challenging point of sensors application in computational imaging is the inverse problem established between sensors acquisition and reconstructions. Traditional physical retrieval methods with certain sensors applications usually result in poor reconstruction due to the highly ill-posedness, particularly in moving object imaging. Thanks to the development of neural networks, data-driven methods have greatly improved its accuracy, however, heavy reliance on data volume has put great pressure on data collection and dataset fabrication. To the best of our knowledge, we firstly propose a sensor-dominated restoration prototype termed “event-enhanced passive NLOS imaging prototype for moving objects with physical embedding” (EPNP), which illustrated the application of dynamic vision sensors in NLOS imaging. EPNP induces an event camera for feature extraction of dynamic diffusion spot and leverages simulation dataset to pre-train the physical embedded model before fine-tuning with limited real-shot data. The proposed EPNP prototype is verified by simulation and real-world experiments, while the comparisons of data paradigms also validate the superiority of event-based sensor applications in passive NLOS imaging for moving objects and perspectives in advanced imaging techniques.
\end{abstract}

\begin{IEEEkeywords}
Computational imaging, Dynamic vision sensor, Event Camera, Feature Representation, Non-line-of-sight imaging.
\end{IEEEkeywords}

%
\IEEEpeerreviewmaketitle

\section{Introduction}
%
%
%
%

\IEEEPARstart{N}{on-line-of-sight} (NLOS) imaging, as a detection and sensing technique for hidden objects outside the direct line-of-sight, has shown promising potential for applications in medical imaging, autonomous driving, reconnaissance and rescues \cite{Kirmani:1,Maeda:2,Faccio:3}. In modern applications, these corner imaging techniques have raised increasing demands for high sensitivity and rapid dynamic sensing ability. Fortunately, with the adventure of novel sensors and its implementation in computational imaging, NLOS imaging has gradually moved towards practical application with simplified and intelligent systems.

In some review articles \cite{Faccio:3}, \cite{Geng:4},  the application of NLOS imaging technique can be classified into active and passive NLOS imaging, depending on whether a controllable light source is used for sensors acquisition. Active NLOS imaging \cite{Geng:4} primarily adopts a pulsed laser as active probing source and acquires three-bounce photon information to reconstruct hidden targets. In passive NLOS imaging \cite{Geng:4}, diffuse reflection on the relay wall is captured and analyzed by data mining with physical priors. From the perspective of sensor acquisition in computational imaging, different dimensions of light field are captured for subsequent information interpretation \cite{Hang:5}. Therefore, the forward models of these two types of NLOS imaging with different sensors applications are completely different.

Thanks to the prosperous evolution of novel sensors and imaging devices, we have witnessed rapid development in NLOS imaging over the past decade. Active NLOS imaging mainly leverage the high sensitivity of single photon avalanche diodes (SPAD) in single-photon detection and counting \cite{Piron:6}, and achieve high temporal resolution for ultra-fast photography and probing \cite{Velten:7}. Generally, according to whether the time parameter of light field transportation is sampled at sub-picosecond scales by the camera, imaging technique can be classified as transient imaging \cite{Shen:8,Feng:9} and steady imaging \cite{Chen:10}. Most active NLOS imaging employs transient information and realizes depth reconstructions for occluded targets by decoding the time-of-flight (TOF) information adopted by the “three-bounce” photon-counting histogram \cite{Velten:11} obtained by SPAD camera. Different algorithms have been proposed to interpret the NLOS light field \cite{Liu:12} captured by different imaging sensors, such as back projection (BP) \cite{Jin:13}, light-cone transform (LCT) \cite{OToole:14}, compressed sensing (CS) \cite{Ye:15}, and fk-migration \cite{Lindell:16} for photon-counting data processing, which show satisfactory reconstruction results for static NLOS objects. Recently, SPAD arrays are also used for active NLOS imaging to further increase the spatial resolution of single photon detection \cite{Riccardo:17}. However, active NLOS system requires precise calibration \cite{Pan:18, Wang:19} and performs repeated scanning, which is time consuming and easy to suffer from perturbation. Moreover, light field interpretation requires sophisticated algorithms and significant computational resources, making real-time applications \cite{Nam:20} challenging and potentially limiting the practicality in certain scenarios especially dynamic scenes. 

In contrast, passive NLOS imaging mainly falls in the field of steady imaging \cite{Chen:10}, which uses sensors with passive detecting paradigm to capture the directly diffused light field for reconstructions. Passive NLOS imaging has shown promising application potential owing to its simple system and efficient data acquisition in ordinary imaging circumstances \cite{Metzler:21}. In this paper, we mainly focus on this type of simplified and environmentally flexible imaging method, which holds much promise for sensor-dominated applications in computational imaging. Specifically, in steady-state imaging, a commonly used pipeline is to employ a flat-panel photonic sensor to capture the diffuse reflection information on the relay surface \cite{Beckus:22}, and reconstruct the hidden object with frame-based intensity information \cite{Wang:23}. With the development of computational imaging \cite{Mait:24} and passive NLOS sensing, different dimensions of diffuse reflection observations are proposed to handle certain scenarios, such as phase retrieval \cite{Metzler:21, Katz:25} based on traditional intensity structures \cite{Beckus:22,Wang:23,Katz:25,Saunders:26}, with frame camera, polarized ques \cite{Tanaka:27} with polarization sensors, optical TOF techniques \cite{Boger:28} with TOF sensors, thermal methods \cite{Maeda:29} with infrared sensors and spectral content approaches \cite{Hashemi:30} with spectral sensors. These methods require different imaging sensors for information acquisition. However, limited by the camera shutter speed, these types of frame-based information of different light field dimensions cannot be temporally resolved, which implies that different depths and textures of object information are interleaved within the integrated time of each snapshot. In particular, for NLOS moving objects imaging \cite{Wang:19, Gariepy:31, Chan:32, He:33}, the frame-based passive NLOS detection mode suffers more from serious degradation on the relay surface and superposition disturbance from isotropic diffuse reflection in close pixels. Towards this end, we introduce event camera \cite{Gallego:34} a novel bio-inspired sensor, for passive NLOS sensing in dynamic scenes, which captures spatio-temporal information with its asynchronous sampling paradigm. Consequently, we can establish an inverse problem based on the single reflection forward model, where optimization algorithms are prompted \cite{Mu:35} for a better solution.

For the applications of sensors acquisition in passive NLOS imaging, we trace back to the original intention of computational imaging, which implements sensor-dominated physical embedding to improve the reconstruction performance. \cite{Feigin:36}. The essence of computational imaging is to retrieve information or features with limited projections of the physical light field captured by the sensor \cite{Ozcan:37}. Since original frame-based intensity data act as low dimensional acquisitions, the condition number of the ill-posed problem is enlarged. Therefore, current research on passive NLOS mainly focuses on reconstructing the object shape or positioning by mining the diffuse information on the relay surface and optimizing the blind deconvolution using data-driven approaches \cite{Zheng:38,Li:39,Cao:40}. Moreover, without the guidance from physical mechanisms, the generalization ability of the deep learning model depends heavily on data diversity and volume. Consequently, some physical embedded data-driven methods \cite{Wu:41, Chen:42} for sensors data utilization have emerged to alleviate this dilemma, however, the demand for data volume imposes significant pressure on data collection and dataset fabrication. Nevertheless, these end-to-end deep learning approaches \cite{Chen:10,Ye:15} only perform well in static NLOS imaging \cite{Lindell:16}, but show shortcomings in moving target reconstructions due to the superposition of dynamic diffuse reflection. This makes the application of frame-based sensors challenging in real-world environment with dynamic scenes or moving objects. Leveraging the sensitivity of event-based sensor in dynamic vision and feature extraction \cite{Wang:43}, we establish an event-based and physics-informed passive NLOS imaging framework for moving objects to put forward its applications in practical scenarios. 

In this work, we propose a passive NLOS computational imaging framework with the enhancement of event-based vision, termed “Event-enhanced Passive Non-line-of-sight imaging for moving objects with Physical embedding” (EPNP), which is an application of event-based sensors and improves the NLOS imaging performance of moving objects with the physical priories supported by simulation datasets. In EPNP prototype, we train the end-to-end(E2E) network on large-scale sensor-dominated simulation datasets and fine-tune the model with few real-shot datasets. With the instruction of the pre-trained model, more dynamic light field propagation parameters and sensor acquisition characteristics are included, which embeds physical transportation mechanics and priors to improve the reconstruction ability and efficiency. The main contributions of this study can be summarized as follows.

\begin{itemize}
\item The introduction of event-based vision to NLOS imaging enables in-sensor computing for sensor-dominated feature extraction in dynamic light field, opening up new possibilities for novel sensors data acquisition in dynamic NLOS imaging and computational imaging.
\item The physical constrained data-driven method under EPNP framework provides a new gateway for passive NLOS imaging with novel sensors data utilization and limited real-world dataset in practical scenarios.
\item The fusion of event-based data with frame-based data further improves the reconstruction performance of dynamic scenes in unconventional imaging and unlocks the ability and broaden the application extent of intelligent sensors.
\end{itemize}

\section{Principles and Methodology}

In this section, we introduce the data acquisition principle of event-based sensors and the event-paradigm forward model in passive NLOS for physical embedding and further simulation prototype establishment.

\subsection{Event-based Vision}

Event cameras, also known as bio-inspired neuromorphic vision sensors, are a type of imaging and sensing devices that response significantly different from traditional frame-based cameras \cite{Gallego:34}. Instead of capturing the absolute brightness of the full frame at fixed intervals, event cameras detect and report per-pixel brightness changes asynchronously and in real-time. The novel sampling paradigm brings extraordinary sensitivity to dynamics for event cameras, which results in high temporal resolution, high dynamic range, and low latency. Therefore, event-based vision has significantly broadened various applications in challenging scenarios for traditional cameras, encompassing diverse applications in high dynamic range high speed imaging \cite{Rebecq:44}, robotics, autonomous vehicles \cite{Xu:45} and object detection \cite{Li:46} in sophisticated optical field.

\begin{figure}[ht!]
\centering
\includegraphics[width=\linewidth]{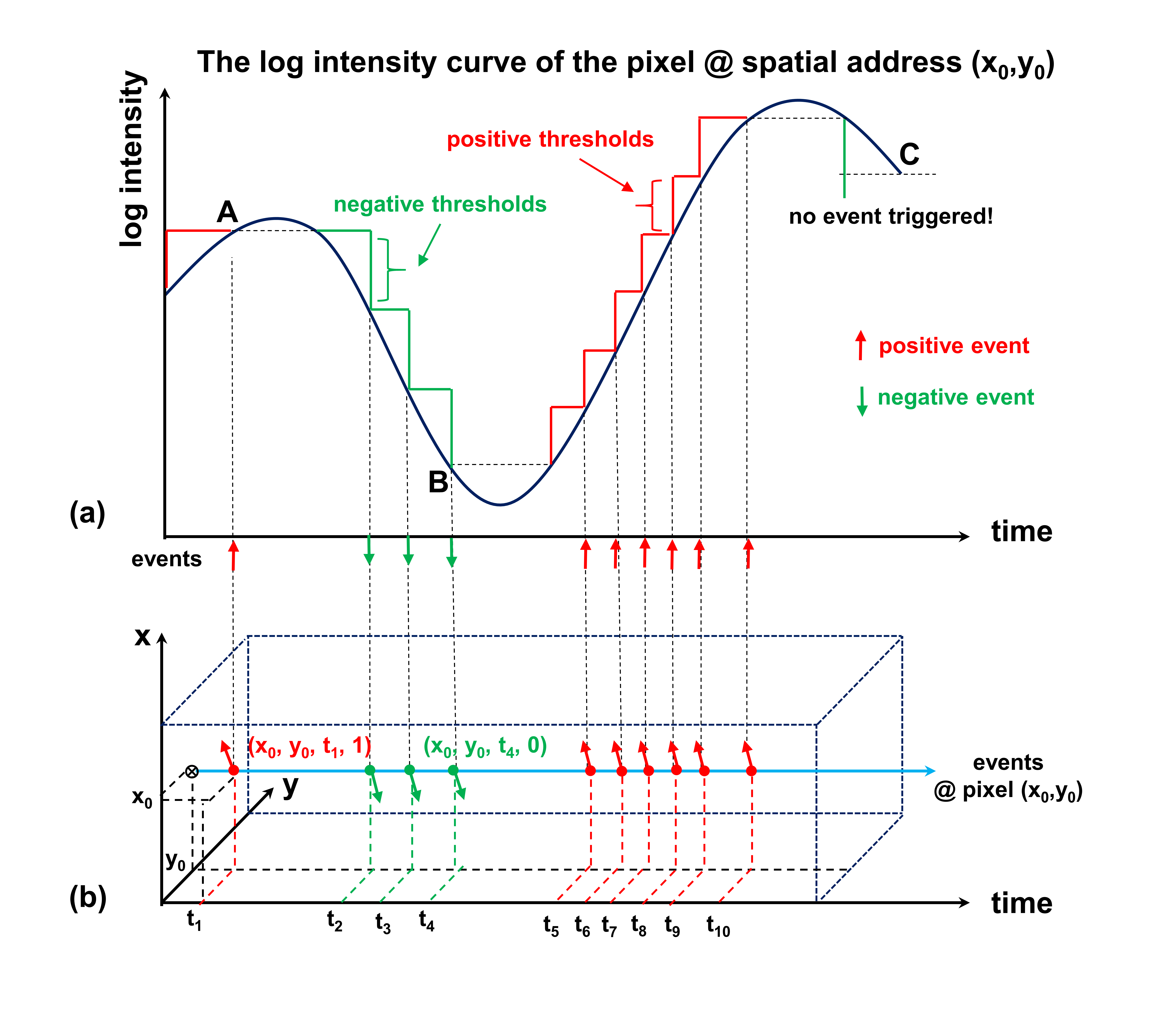}
\caption{Illustration of the working principle of event camera at a specific pixel.(a) shows the log form intensity curve of the pixel at spatial address $(x_0,y_0)$, (b) demos the triggered events in sptio-temporal cube representation.}
\label{fig:event}
\end{figure}

The sampling principle of an event camera is illustrated in Figure \ref{fig:event}. When the pixel-wise logarithmic intensity of brightness changes exceeds the pre-set threshold, the specific pixel is triggered and an event is recorded binomially, either positive or negative. Each pixel continuously monitors the logarithmic form of the intensity, awaiting sufficient amplitude changes to fire another event. 

As shown in Figure \ref{fig:event}(a), a positive event $(x_0, y_0, t_1, 1)$ is triggered at point A and a negative event $(x_0, y_0, t_4, 0)$ at point B, whereas no event is triggered at point C since the change in intensity does not reach the pre-set threshold controlled by the bias voltage. Additionally, the 4-dimensional event-based data can be represented by a set of sparse scatter points in the spatio-temporal cube in Figure \ref{fig:event}(b), which intuitively expresses the property of event excitation. One can see from Figure 1, an event is characterized by four distinct parameters: spatial address $x$, $y$, time stamp $t$, and polarity $p$, which is noted as $\epsilon_i$ for the $i^{th}$ event,

\begin{equation}
\epsilon_i = \{x_i, y_i, t_i, p_i  \},
\label{eq:event}
\end{equation}

\noindent where $x$ and $y$ denote the spatial address of the fired pixel, $t$ records the time stamp of the triggered moment, $p$ represents the binary polarity flag indicating an increase or decrease in brightness, respectively.

\subsection{Forward Model of Passive NLOS imaging}
The forward model of passive NLOS imaging \cite{Saunders:26, Geng:47} plays a crucial role in explaining how light is transported in hidden scenes and how it is captured and interpreted by the sensor for reconstruction. It forms the physical foundation for developing algorithms and techniques to improve the reconstruction quality of passive NLOS imaging.

\begin{figure}[htbp]
\centering
\includegraphics[width=\linewidth]{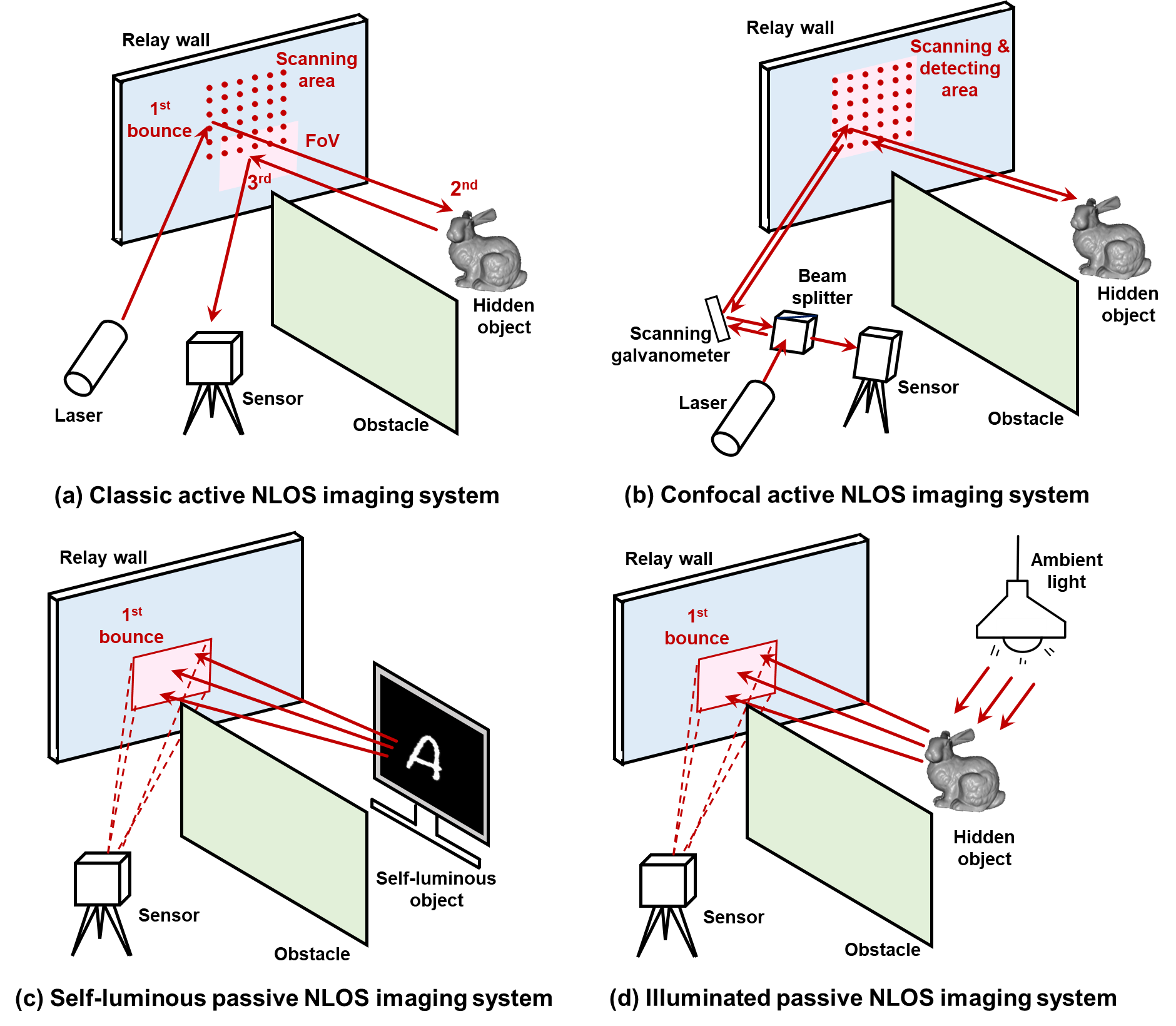}
\caption{Classification of NLOS imaging system according to the experimental setting.}
\label{fig:class}
\end{figure}

From the perspective of experimental settings, typical NLOS imaging system can be classified into four categories, as shown in Figure \ref{fig:class}.

According to whether the detection focal point is coincident with the scanning point, active NLOS imaging is classified into classic system with laser illumination and confocal system with scanning galvanometer, as shown in Figure \ref{fig:class} (a) and (b) respectively. However, in passive NLOS imaging, a conventional way to classify the imaging system is according to the illumination type of the hidden object. As shown in Figure \ref{fig:class} (c) and (d), it can be subdivided into self-luminous and illuminated situation.

\begin{figure}[htbp]
\centering
\includegraphics[width=0.7\linewidth]{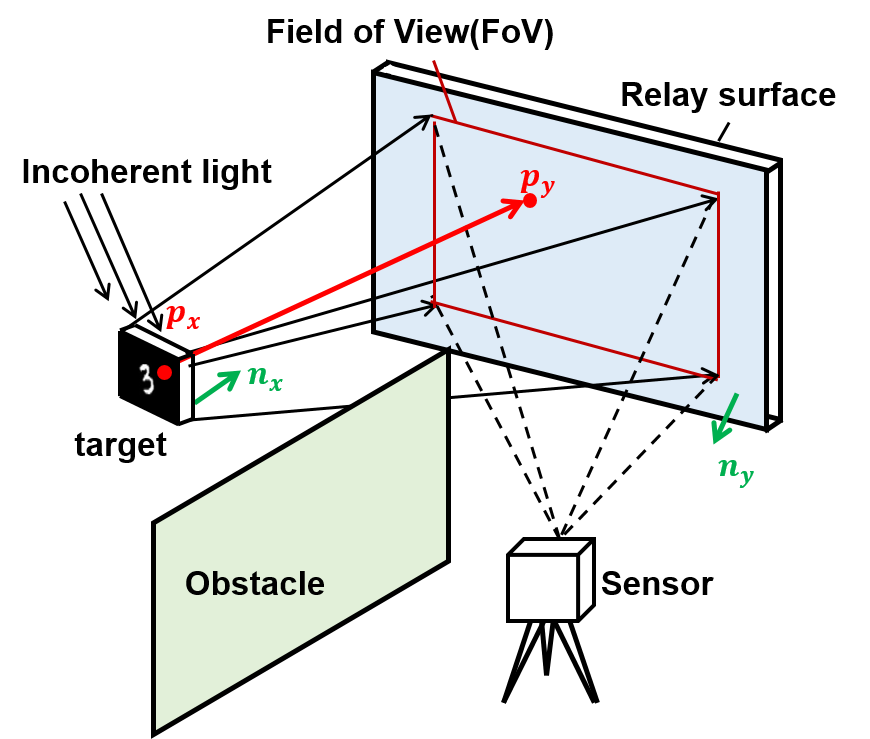}
\caption{Symbolic description of the forward model in passive NLOS settings.}
\label{fig:fm}
\end{figure}

We assume the surface of the illuminated object as secondary light source when ambient light is steady and invariant, so that both of the two passive NLOS situations can be considered as several point source array which discretize the hidden object for modelling. In both of these settings, the sensor captures the diffusely reflection light field on the relay wall. Therefore, in particular passive NLOS setups, we express the forward model by describing the light field mapping relation between the hidden object and the diffusely reflecting surface, which is known as relay surface. Specifically, as shown in Figure \ref{fig:fm}, the irradiance of a patch at $p_y$ on the relay surface \cite{Saunders:26} can be expressed as

\begin{equation}
I \left( p_y \right) = \iint_{p_x \in F}{A_\measuredangle \left( p_x, p_y \right) \cdot I \left( p_x \right) dp_x +B},
\label{eq:fm1}
\end{equation}

\noindent where $I \left( p_x \right)$ denotes the source irradiance emitted from the hidden target, $I \left( p_y \right)$ denotes the irradiance distribution on the relay surface, $F$ stands for the whole source pixels corresponding to the detecting field of view (FoV) and $B$ represents the background noise term including ambient light noise and detection noise. Taking the geometric relationships of light field propagation into consideration, $A_\measuredangle \left( p_f , p_y \right)$ explicates the point-to-point transmission weighting for different position relationships \cite{Saunders:26}, which is specified as

\begin{multline}
A_\measuredangle \left( p_x , p_y \right) = \frac{ cos \left[ \measuredangle {\left( \vec{p_y} - \vec{p_x} , \vec{n_x}\right) } \right] \cdot cos \left[ \measuredangle { \left( \vec{p_x} - \vec{p_y} , \vec{n_y}\right)}\right]}{\left | \left | {\vec{p_y} - \vec{p_x}}\right | \right |_2^2} \\ \cdot \mu \left( p_x, p_y \right),
\label{eq:fm2}
\end{multline}

\noindent where $\measuredangle{\left( \vec{p_y} - \vec{p_x} , \vec{n_x}\right)}$ and $\measuredangle{ \left( \vec{p_x} - \vec{p_y} , \vec{n_y}\right)}$ depict the angle between the transmission vector $  \vec{p_y} - \vec{p_x} $ formed by mapped pixel pair $ \left( \bold{p_x} , \bold{p_y} \right) $ and hidden object with the normal vector on each of their surface plane, respectively, as shown in Figure \ref{fig:fm} (a). The coefficient $\mu$ represents the bidirectional reflectance distribution function (BRDF), parameterizing the reflective properties of different areas on the relay surface, which is regarded as a constant in this study when the relay surface can be approximately modeled as an isotropic diffusely reflecting surface. The cosine form of these two angles endures weights for optical transmission, reflecting the relative brightness and visual angle of different points in the scene and at different positions on the relay surface. In addition, the Euclidean norm in the denominator is utilized to characterize the attenuation with range.

When substituting Eq. \ref{eq:fm2} into Eq. \ref{eq:fm1}, we summarize the forward model of passive NLOS imaging, and establish a simplified discrete version of the detection function on the relay surface which outlines the physical mechanism and suitable for further simulation at the same time.  

\begin{equation}
I \left( p_y \right) = h \left( p_x, p_y \right) \ast I \left( p_x \right) +B,
\label{eq:fm_conv}
\end{equation}

\noindent where h denotes the point spread function (PSF) of the entire transmission system. However, h is difficult to express analytically by linear functions directly because the propagation process and the diffuse reflection on the relay surface make the system a spatially variant system. Furthermore, when detecting and imaging NLOS moving objects with passive imaging sensors, the light field information will also superpose in time domain within the sensor's integration time, increasing the ill-posedness and creating more difficulties for the simulation.

Fortunately, the pixel-wise convolution form is similar to the diffraction of light field, where each pixel on the detection plane interacts with each equivalent light source pixel on the object plane.
Inspired by the wave-based method proposed by X. Liu et al. \cite{Liu:12}, we assume the illuminated target as several secondary point sources and the detecting FoV on the relay surface as virtual sensor array so that the pixel to pixel transmission process can be split into Rayleigh-Sommefeld Diffraction (RSD) propagation and interaction with the relay surface.

To simplify the modeling and simulation process of the sensors data acquisition, we use the angular spectrum form of RSD to model and calculate the propagation process, which supplies the linear portion. Then, we adopt random scattering theory to demonstrate the diffusely reflection, conducting the non linearity and interactional effects. These two sequential processes are expressed by PSF and transmission function (TF), as shown in Eq. \ref{eq:fm_conv2}.

\begin{equation}
I \left( p_y \right) = \Phi \left( h_{RSD} \left( p_x, p_y \right) \ast I \left( p_x \right) \right) + B,
\label{eq:fm_conv2}
\end{equation}

\noindent where $h_{RSD}$ represents the PSF of RSD, and $\Phi \left( \cdot \right)$ represents the interaction function with the relay surface. Specifically, a simplified discrete version of Eq. \ref{eq:fm_conv2} can be written in matrix form as:

\begin{equation}
y = S \cdot D \cdot x + B = A \cdot x + B ,
\label{eq:fm_conv3}
\end{equation}

\noindent where D and S act as the transmission matrix of RSD and diffuse reflection, respectively, $ A = S \cdot D $, $y$ denotes the measurement and $x$ stands for the hidden scene.

Usually, based on the forward model, the corresponding inverse problem can be transformed into an optimization problem for restoration, i.e. the utilization process of sensors data,

\begin{equation}
\hat{x} = arg \min\limits_{x} \| \hat{A}x-y \|_2^2 + R(x) ,
\label{eq:fm_conv4}
\end{equation}

\noindent where $R(x)$ is the regularizer.

\subsection{Event paradigm representations and reconstructions in passive NLOS}

With the superiority in dynamic information sensing, event cameras are widely used in robotics, computational photography, and vision tasks especially in challenging environments. 
In this study, we introduce an event representation method to excavate event-based data and extract spatio-temporal features from sensor-dominated dynamic diffusion spot acquisitions, which contains efficient information of the target movement.

Compared with traditional direct-view imaging, when processing NLOS dynamic scenes, the movement of hidden targets raises additional spatio-temporal aliasing for passive NLOS imaging, increasing the blindness and difficulty of inverse restoration. The challenge of passive NLOS imaging for moving objects with imaging sensors is the adequate and accurate acquisition of dynamic diffusely reflected information without distortion, and efficient dynamic features extraction from the captures for better reconstruction. In this work, an event-based sensor is proposed to mitigate the aliasing of dynamic light field and extract more efficient movement information as well as the spatio-temporal texture feature of the diffusion spot on the relay surface. 
However, the sparse asynchronous paradigm of event data is not applicable for direct substitution in the forward model of passive NLOS. Therefore, event-based sensors data representations must be implemented for information dimension conversion and a more efficient dynamic feature expression. Different representations of the event data are established for corresponding applications, for example, the Sparse Event Tensor (SET) \cite{Guo:48} for vision transformer application in object detection, single pixel event tensor \cite{Zong:49} for applications in image reconstruction, and Event Graph \cite{Li:50} for spatial graph convolution application in computer vision tasks.

\begin{figure}[htbp]
\centering
\includegraphics[width=\linewidth]{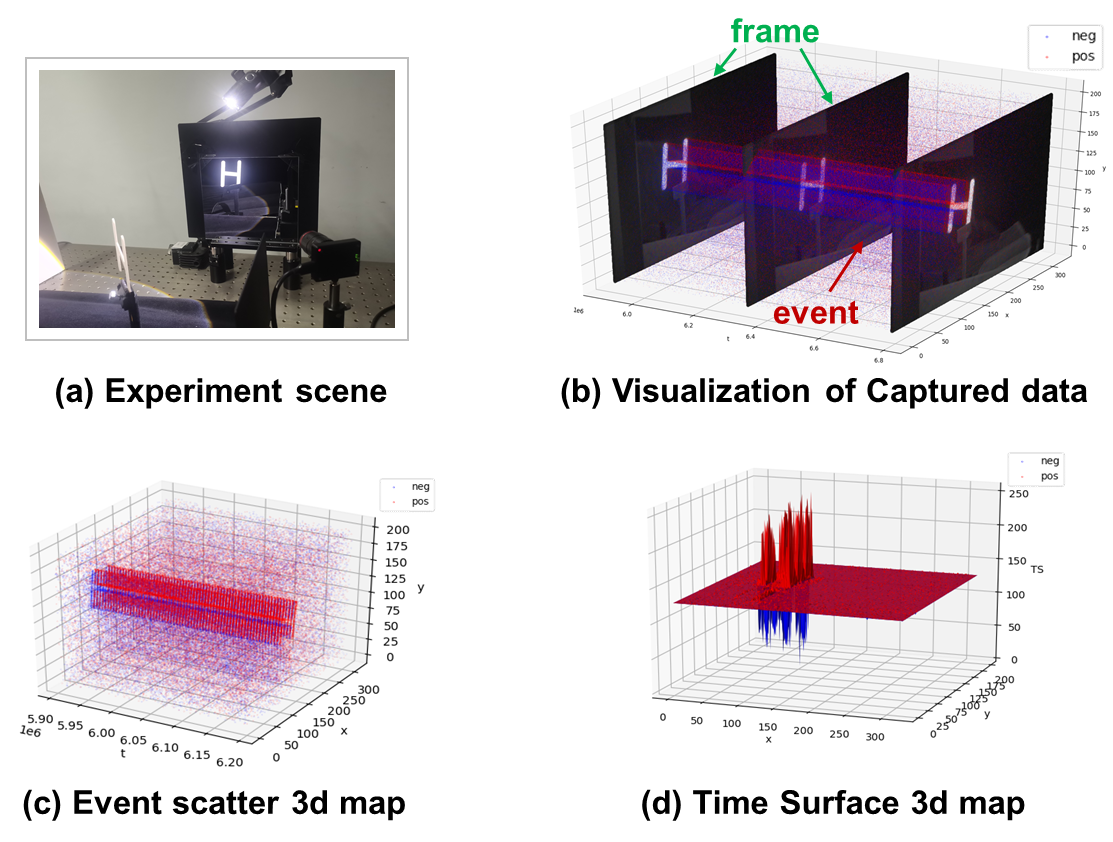}
\caption{Representation and visualization of event-based and frame-based data in passive NLOS scenarios.}
\label{fig:event-re}
\end{figure}

We demonstrate the event-based representation \cite{Wang:51,Lagorce:52} we used by visualizing a set of real-shot event data. For an intuitive illustration, we temporarily use a mirror as the relay surface, and the hidden target “H” moves from left to right in the sensor's FoV.
The difference in data format between event-based and frame-based data is visualized in Figure \ref{fig:event-re} (b), where the event-based paradigm can record inter-frame dynamic information. As shown in Figure \ref{fig:event-re} (c), the asynchronous captured events are represented by scatter points in 3-dimensional cuboid, representing the distinct parameters $x$, $y$, and $t$, whereas the color represents the polarity. To extract spatial and temporal feature from event data and convert into a suitable format for the forward model, we perform Time-surface (TS) calculation on the selected voxel grid \cite{Lagorce:52} and obtain the TS value according to the polarity, as shown in Figure \ref{fig:event-re} (d). The TS value expresses both temporal and spatial correlations of events in the time interval of the selected voxel grid, which can be specified as:

\begin{equation}
S_i \left( \rho ; p \right) = e^{- \left[ t_i - t_i^* \left( \rho ; p \right) \right] / \tau} ,
\label{eq:ts1}
\end{equation}

\noindent where $S_i$ denotes the TS value of event $e_i$, the exponential kernel supplies the decay of spatial and temporal relevance with other event points within the spatial neighborhood $\rho$. The normalized context time stamp $t_i^* \left( \rho ; p \right)$, represents the time-context standard for all events whose spatial address falls in the neighborhood within radius $\rho$ from the incoming $e_i$, defined as Eq. \ref{eq:ts2}.

\begin{equation}
t_i^* \left( \rho ; p \right) = \max\limits_{j \leqslant i} \{ t_j | r_j \in r_i + \rho, p_i = p \} ,
\label{eq:ts2}
\end{equation}

\noindent where $r_j$ and $r_i$ denotes the spatial address vectors of event $e_i$ and $e_j$. $t_i^*$ plays an important role in determining the temporal center of the event points involved in calculation. $p$, $\tau$ are the polarity and time constant, respectively.

The forward model of NLOS imaging structures the light field transmission and the information acquisition process of the sensor. However, the sparse and asynchronous event-based data cannot be directly substituted into the steady state NLOS forward model. Therefore, data processing of these novel sensors information paradigm is performed to achieve conversions in information dimension and extract more efficient dynamic features with event-based representation.

With the representation of event-based data, we establish the event form forward model of passive NLOS based on our previous study \cite{Wang:43}. In addition, to provide a matrix-like mathematical expression, we divide the event data into several temporal voxel grids \cite{Lagorce:52} and calculate the TS map on each voxel grid to substitute the measurement matrix in Eq. \ref{eq:fm_conv3}. By defining the element-wise intensity of pixel $i$ at the center timestamps $t_i$ and $t_i’$ of two adjacent voxel grids as $I_{ y,i}$ and $I_{y,i}'$, the event form detection function can be written by the pixel-wise intensity difference as:

\begin{multline}
\lg I_{y,i}'- \lg I_{y,i} = \sum\limits_{i} A_i \left( I_{x,i}'- I_{x,i} \right) + B' - B
\\  = A_i \lim\limits_{\Delta t \rightarrow t_0} \Delta x^m_i + B_s ,
\label{eq:efm1}
\end{multline}

\noindent where $t_0$ denotes the time interval of a voxel grid, and the limitation form explicit the accumulation of target intensity changes $I_{x,i}'- I_{x,i}$ at each pixel, annotated as $x^m_i$, which contains the movement information of the target during interval $t_0$. 
In addition, the sampling paradigm of event cameras can effectively eliminate ambient noise so that only sensor noise remains, $B_s = B'- B$.

\begin{figure}[htbp]
\centering
\includegraphics[width=\linewidth]{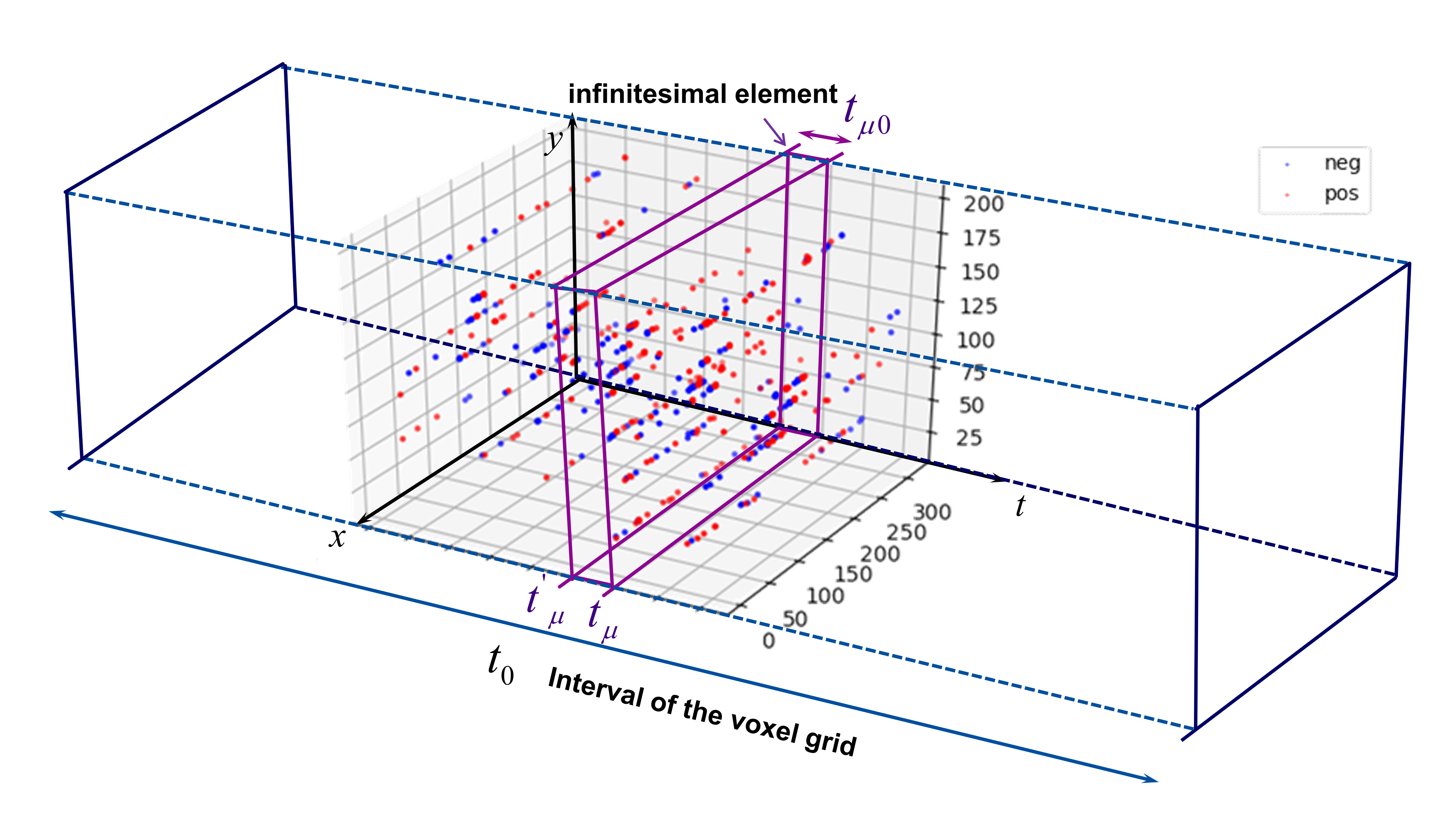}
\caption{Illustrations of infinitesimal element bin and voxel grid for event-based data.}
\label{fig:voxel}
\end{figure}

We can then rewrite the forward model by analytically substituting TS calculation into the detection function as Eq. \ref{eq:fm_conv3}. Before performing TS calculation to obtain the featured map, we define a discrimination function $\delta$ to indicate whether an event is triggered. It is worth noting that events are triggered pixel by pixel instantaneously, so the time interval of the voxel grid needs to be decomposed to microsecond level, which is the temporal limitation of general event-based sensors. As shown in Fig. \ref{fig:voxel}, we split the voxel gird into several infinitesimal element bins with time interval $t_{\mu 0}$. Suppose $I_{y,i}^{t_{\mu}’}$ and $ I_{y,i}^{t_{\mu}}$ denote the intensity of pixel $i$ at the initial timestamp $ t_{\mu}’$ and the end timestamp $ t_{\mu}$, respectively. The relationship between voxel grid and infinitesimal element bin is also illustrated in Fig. \ref{fig:voxel}, where the purple box represents the infinitesimal element bin, ensuing that at most one event is triggered for each pixel in each bin.

According to the detected intensity change $\lg I_{y,i}^{t_{\mu}’} - \lg I_{y,i}^{t_{\mu}}$ within $t_{\mu 0}$, the discrimination function $\delta$ determines whether an event is triggered for the pixel.

\begin{equation}
\delta \left(\lg I_{y,i}^{t_\mu '} - \lg I_{y,i}^{t_\mu} \right) = \left \{ \begin{array}{rcl} 
\left[x_i, y_i, t_i, 1 \right] & , & \lg I_{y,i}^{t_\mu '} - \lg I_{y,i}^{t_\mu} \geq C_+ ,   \\
\left[x_i, y_i, t_i, 0 \right] & , & \lg I_{y,i}^{t_\mu '} - \lg I_{y,i}^{t_\mu} \geq - C_-  \\
null & , & -C_- \textless \lg I_{y,i}^{t_\mu '} - \lg I_{y,i}^{t_\mu} \textless C_+ \end{array} \right.
\label{eq:ds_event}
\end{equation}

\noindent where $C_+$ and $ C_-$ are the preset triggered threshold for positive event and negative event respectively.

Then, TS calculation is performed in a set of voxel grids, the context information of movements is accumulated, and the spatio-temporal features of the moving target are concised into a matrix $E_{map}$,

\begin{equation}
E_{map} = \sum\limits_{y \in FoV} S \left[ \sum\limits_{ \Delta t \rightarrow t_0, \forall i} \delta \left( \lg I_{y,i}' - \lg I_{y,i}  \right) \right] ,
\label{eq:tsmap}
\end{equation}

\noindent where $S$ denotes TS calculation, and all triggered events in the FoV on the relay surface contribute to the $E_{map}$. The 2d matrix form is the projection of TS 3d map on $xoy$ plane, in which the passive map and negative map are normalized in one matrix. Therefore, the event form forward model is written as,

\begin{equation}
E_{map} = A \cdot x^m + B_s.
\label{eq:event_fm1}
\end{equation}

According to the classic reconstruction methods from event data to images \cite{Rebecq:44}, the relationship between the absolute intensity and target movement conveyed by the TS map can be bridged as an implicit function, i.e. $x^m = \Gamma \left( x \right)$. Thus, Eq. \ref{eq:event_fm1} is converted to a function with respect to the hidden scene x,

\begin{equation}
E_{map} = A^* \cdot x + B_s ,
\label{eq:event_fm2}
\end{equation}

\noindent where $A^*$ is the adapted transmission matrix, with a relatively large condition number, i.e. $A^* \cdot x = A \cdot x^m$. Typically, the reconstruction problem of event-based NLOS is transformed into solving the inverse problem described by Eq. \ref{eq:event_fm2}, which retrievals the hidden target by solving the optimization problem,

\begin{equation}
\hat{x} = arg \min\limits_{x} \| \hat{A^*}x-y \|_2^2 + J ,
\label{eq:event_fm3}
\end{equation}

\noindent where J denotes the priori induced by the physical transmission process and event-based feature representation.

\section{Event-enhanced Passive Non-line-of-sight imaging for moving objects with Physical embedding}

In this section, we establish the reconstruction pipeline for sensors data utilization. Specifically, we propose a novel framework for enhancing passive NLOS imaging of moving objects, which compensates for the shortcomings of end-to-end (E2E) reconstructions. The event-based detecting paradigm enhances the perception capabilities of the dynamic light field, whereas the simulated pre-training method is used to better illustrate and embed the physical forward transmission characteristics.

\subsection{EPNP Prototype}

Recall that the spatio-temporal feature of the hidden target movements can be expressed by the TS map of diffusion spot movements on the relay surface, and the event paradigm NLOS forward model is consistent with the traditional frame-based one, we performed an E2E deep learning method to optimize the retrieval process in our previous work \cite{Wang:43}. To further improve the reconstruction quality, we need to overcome the over-fitting and strong reliance on data distribution introduced by data-driven approaches. However, the data acquisition process for illuminated objects makes it difficult to fabricate data sets with large data volumes and rich diversity.

To the best of our knowledge, we firstly propose the "Event-enhanced Passive Non-line-of-sight imaging for moving objects with Physical embedding" (EPNP) prototype to establish large-scale simulation datasets for pre-training the physical embedded model, and fine-tune with limited real-shot data. In the practical application of event-based sensor in NLOS imaging, the EPNP prototype acts as a Plug and Play (PnP) structure, which combines the shared physical mechanism of forward transmission in passive NLOS. It also allows the adjustment in real-world captures for both frame-based and event-based sensors data.

\begin{figure}[htbp]
\centering
\includegraphics[width=\linewidth]{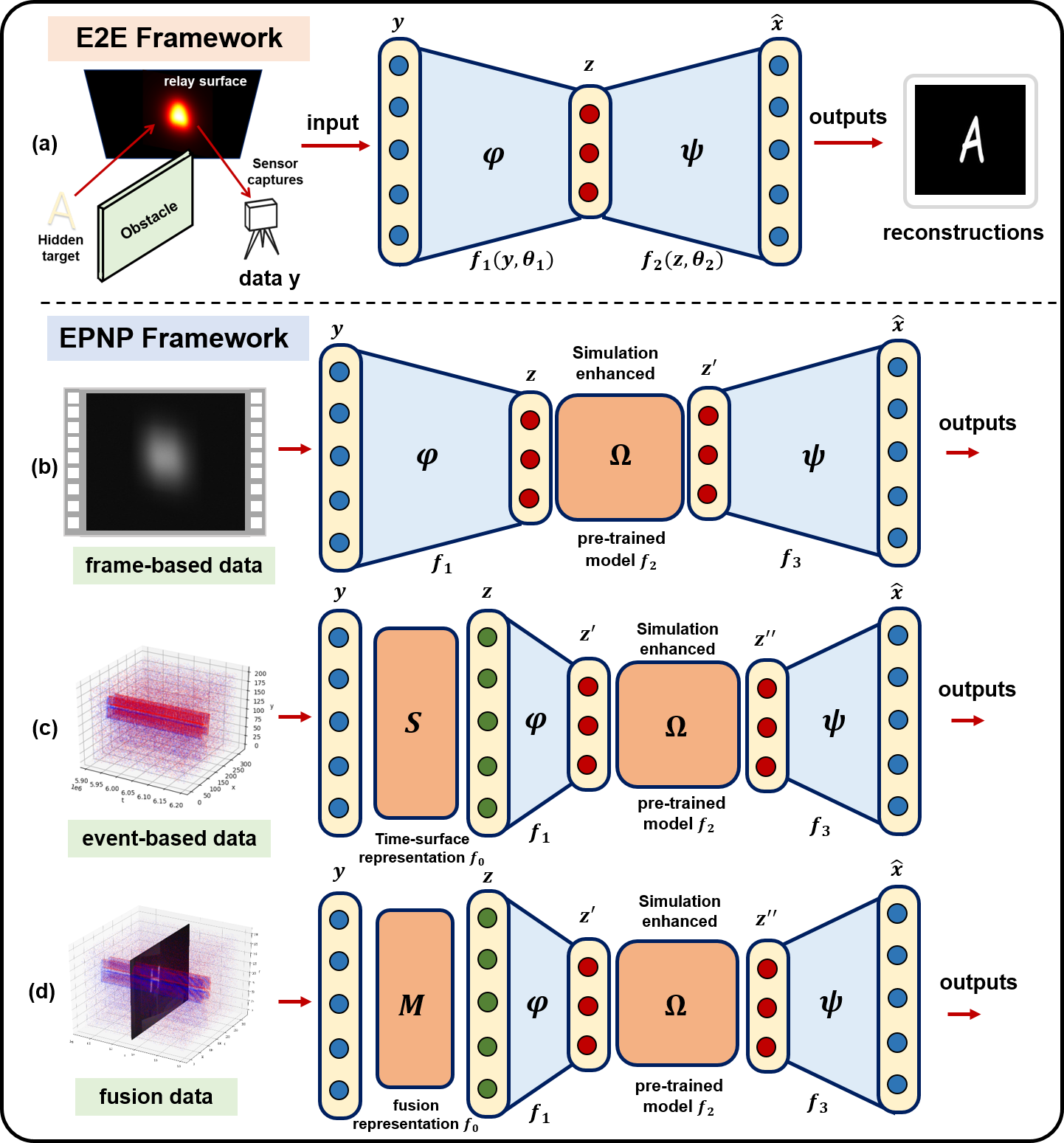}
\caption{Illustration of end-to-end framework and EPNP framework in passive NLOS imaging with different paradigms of sensors data.}
\label{fig:EPNP}
\end{figure}

As shown in Figure \ref{fig:EPNP}, based on the forward model, traditional E2E reconstruction with data-driven method is illustrated in Figure \ref{fig:EPNP} (a), where the sensor captures pass through the encoder and decoder successively for feature extraction in the latent space and reconstructions. The loss function is crafted to optimize the parameters $\theta$ in mapping relations $f_1$ and $f_2$. In this case, the physical fundamentals are implicitly modeled and concealed in data-driven network models. To maximize the effect of physical embedding in guiding the convergence of data-driven network, we propose the EPNP prototype as shown in Figure \ref{fig:EPNP} (b)(c)(d), which adapts different paradigms of the input data: event, frame and fusion forms. The frame-based counterpart embeds the pre-trained simulation-enhanced module $\Omega$ in the latent space to instruct the training when fine-tuning with real-shot data. Similarly, the event-based counterpart adopts an additional TS representation module $S$ to convert the data format. In addition, the fusion form is the combination of a frame-based snapshot with event-based data in voxel grids within a certain time interval, which requires a fusion representation module $M$ before the simulation-enhanced module $\Omega$.

Specifically, we curate the data by simulating large-scale datasets for passive NLOS of moving objects, i.e. the light field simulations of the dynamic diffusion spot intensity information on the relay surface. Then, we train the E2E model with simulation datasets to learn the physical light field transmission mechanism as a pre-trained model. Thus, when limited real-shot datasets are input to the corresponding framework, simulation-enhanced module $\Omega$ supplies physical constraints and priors to alleviate the over-fitting caused by insufficient data diversity and further enhance the reconstruction performance.

\subsection{Simulation Pipeline}

\begin{figure}[htbp]
\centering
\includegraphics[width=\linewidth]{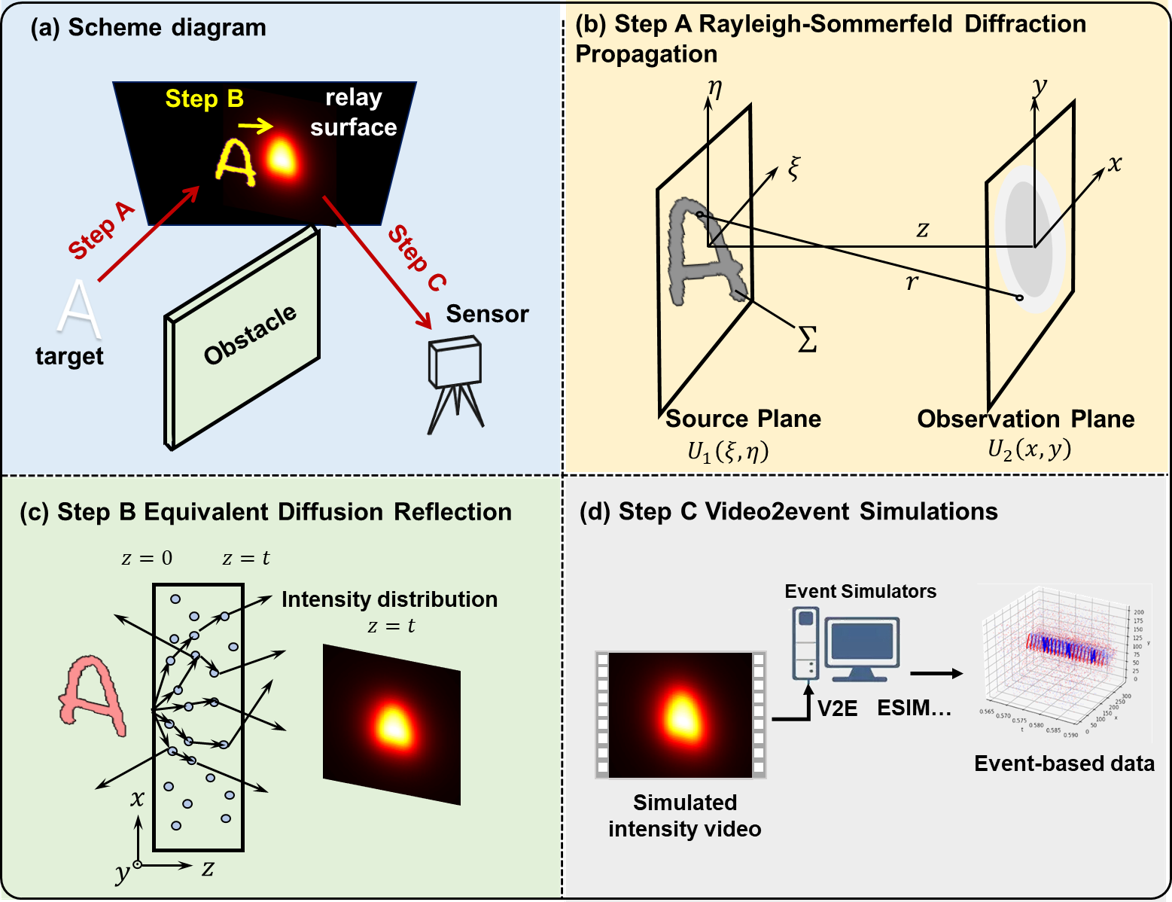}
\caption{The scheme of simulation pipeline in EPNP prototype , (a) is the physical scheme of the simulation, (b) is the diffraction propagation process, (c) is the equivalent diffusely reflection process, (d) is the event generation process.}
\label{fig:simu_pipline}
\end{figure}

The simulation pipeline for passive NLOS imaging of moving objects with event-based sensors in the EPNP prototype can be divided into three main steps as shown in Figure \ref{fig:simu_pipline}.

\subsubsection{Step A: Rayleigh-Sommerfeld Diffraction Propagation}

Recall that the propagation of light field is expressed by discrete convolution in Eq.\ref{eq:fm_conv3}, diffraction integral is an accurate way to describe this process in numerical simulations.

Assuming that the illuminated hidden object can be modeled as a set of secondary point sources, we utilize  Rayleigh-Sommerfeld diffraction (RSD) \cite{Zhang:53} to simulate the forward propagation of the light field emitted from the target before it arrives at the relay surface. Since the temporal phase information is lost with passive NLOS detection, we slightly modify the RSD-based Virtual Wave Optics \cite{Liu:12} to fit the forward model, as shown in Figure \ref{fig:simu_pipline} (b). Suppose the intensity distribution of the hidden object is written as $U_1(\xi, \zeta)$, the light field $U_2(x, y)$ before hitting the relay surface is calculated by Eq. \ref{eq:rsd}.

\begin{equation}
U_2(x,y,z) = \frac{1}{j\lambda} \iint_{-\infty}^{+\infty}{ U_1(\xi, \zeta) \frac{z}{r} \cdot \frac{\exp(jkr)}{r}} d\xi d\zeta .
\label{eq:rsd}
\end{equation}

To expedite the calculation, we adopt the angular spectrum form of RSD to calculate the light field propagation process. Specifically, we approximate the propagation as plane to plane diffraction, rewrite the integral with planar primitives, and get the expression of complex amplitude as Eq. \ref{eq:rsd_2}.

\begin{equation}
\begin{split}
E(x,y) = \iint_{-\infty}^{+\infty} \mathcal{F} \left[ U_1(\xi, \zeta) \right] \exp \left[ jkz \sqrt{1-\lambda^2 f_\xi^2 - \lambda^2f_\zeta^2} \right]  \\ \exp \left[ j2\pi (f_\xi x+f_\zeta y) \right] df_\xi df_\zeta ,
\end{split}
\label{eq:rsd_2}
\end{equation}

\noindent where $a(f_\xi,f_\zeta) = \mathcal{F}\left[ U_1(\xi, \zeta) \right] $ is the angular spectrum, $f_\xi = \frac{cos \alpha}{\lambda}$ and $f_\zeta = \frac{cos \beta}{\lambda}$ is the spatial frequency, and $\alpha$ and $\beta$ are the angles formed by wave vector $k$ with the corresponding axis. The first exponential factor serves as the transmission function $H$. Finally, we take the norm of matrix $E$ and obtain the intensity distribution of light field before interacting with the relay surface.

\subsubsection{Step B: Equivalent Diffusion Reflection}

For a long time, the modeling of random scattering on relay surface has been a bottleneck in NLOS simulations. In this work, we do not pursue exactly accurate modeling of the diffusely reflection properties of the relay surface, because in practical scenarios, the ambient environment arises random disturbance to the physical mechanism of diffusely reflection. Instead, from the perspective of sensor acquisition, we statistically construct the distribution of the light field after the interaction with relay surface.

Since imaging through scattering layers and imaging around corners shares similar diffraction-limited optical imaging system, the imaging technique can be interchanged to some extent, for example via speckle correlations \cite{Katz:25}. The physical mechanism of imaging through scattering medium can also be applied to imaging occluded targets in NLOS area using the light back-scattered from a diffusive surface. Therefore, with assuming the reflection from the relay surface to be Lambertian, we can use scattering theory to equivalently simulate diffusely reflection in “around the corner” circumstance \cite{Katz:25}.

\begin{figure}[htbp]
\centering
\includegraphics[width=\linewidth]{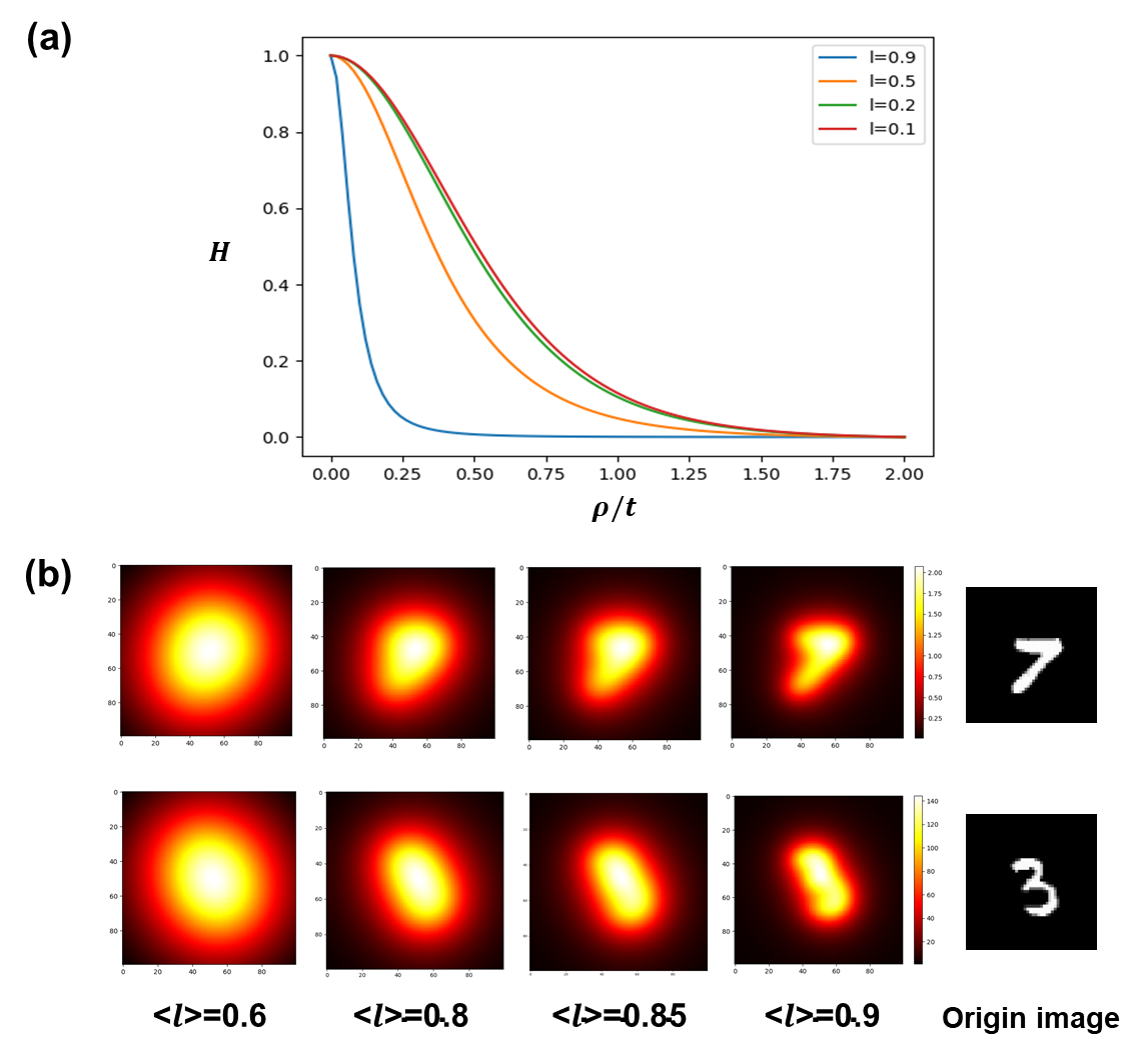}
\caption{Demonstration of parameter $l$ and $\rho/t$ in simulations, (a) tPSF curve, (b) simulation results with different $l$.}
\label{fig:randomwalk}
\end{figure}

As shown in Figure \ref{fig:simu_pipline} (c), we leverage the random walk and diffusion-like model of photons \cite{Gandjbakhche:54} to statistically simulate the photon’s behavior and pattern when interacting with the diffuse surface. Therefore, when recording with photonic sensors, the possibility distribution of photons on the relay surface after interaction can be approximately expressed by the transmission point spread function (tPSF) of turbid media \cite{Rogers:55,Rogers:56} as Eq. \ref{eq:H}.

\begin{equation}
H(\rho) = \frac{\tau}{2\pi} \sum_{k(odd)=-\infty}^{+\infty} \frac{<l> +kt}{ \left[ ( <l> + kt)^2 + \rho^2 \right]^{3/2}} ,
\label{eq:H}
\end{equation}

\noindent where $\rho$ denotes the lateral distance from the start point of random walk, $\tau$ is the ratio of slab thickness $t$ to scattering mean free path $<l>$. We simulate the tPSF curve when the equivalent thickness $t$ is set to unit 1, and normalize the calculation result for demonstration, as shown in Figure \ref{fig:randomwalk} (a). 

The trend of the curve is very similar to that obtained from Lambert's Cosine Law, which verifies the feasibility of simulated diffusion approximation with multiple-path random walk model through turbid media. 

After performing convolution to the incidence image with kernel formed by $H$, the simulation results under different scattering mean free path $<l>$ are shown in Figure \ref{fig:randomwalk} (b). It is obvious that the larger $<l>$ indicates less step in random walk, which results in sharper trend of the curve. The broadening of the curve on the $\rho/t$ axis describes the effect of diffuse dispersion. Compared with the intensity distribution of real-shot image, we select $t=2$, $<l>=0.8$,and $\rho$ sampling from 0 to 2 with 100 discrete points uniformly for simulation.

\subsubsection{Step C: Video2event Simulations}

After simulating of RSD propagation and diffuse reflection, frame-based datasets are ready to be generated when random noise is added. However, to simulate the event-based captures for movement feature extraction, we need to transfer the intensity distribution into event paradigm. As for this paradigm conversion, we generate a video with simulated intensity distribution at different moment through the object movement, and use the start-of-the-art event flow simulation pipeline video2event (v2e) \cite{Hu:57} to simulate the event-based color acquisitions, as shown in Figure \ref{fig:simu_pipline} (d). The v2e simulation is conducted with parameters listed in Table \ref{tab:v2e}.

\begin{table}[htbp]
\centering
\caption{\bf Parameter settings in V2E simulation.}
\begin{tabular}{lcc}
\hline
parameter & value  & symbolic unit \\
\hline
timestamp resolution & $0.001$  & s \\
DVS exposure duration & $0.005$  & s \\
positive threshold & $0.15$  & - \\
negative threshold & $0.15$ & - \\
sigma threshold & $0.03$ & - \\
wight,hight & $(960,600)$ & pixel \\
cutoff hz & $15$ & Hz \\
shot noise rate hz & $0.0015$  & Hz \\
model & SuperSloMo39.ckpt & - \\
\hline
\end{tabular}
  \label{tab:v2e}
\end{table}

\noindent where the timestamp resolution sets the up sampling factor for the input video. It can be combined with the “Slow motion” model that can interpolate frames to ensure DVS events reach at most resolution. The DVS exposure time supplies the fixed accumulation time for event camera to finish the frame event integration. The positive and negative thresholds show the logarithmic form of the intensity change to trigger a corresponding event, and the sigma threshold controls the deviation threshold variation of the logarithmic intensity change. The cutoff frequency and shot noise rate are control parameters of the sensor circuit, where the cutoff frequency is responsible for photoreceptor infinite impulse response low-pass filter, and the shot noise rate determines the temporal noise rate of ON and OFF events.

\subsection{Reconstructions with EPNP}

\begin{figure*}[htbp]
\centering
\includegraphics[width=0.75\linewidth]{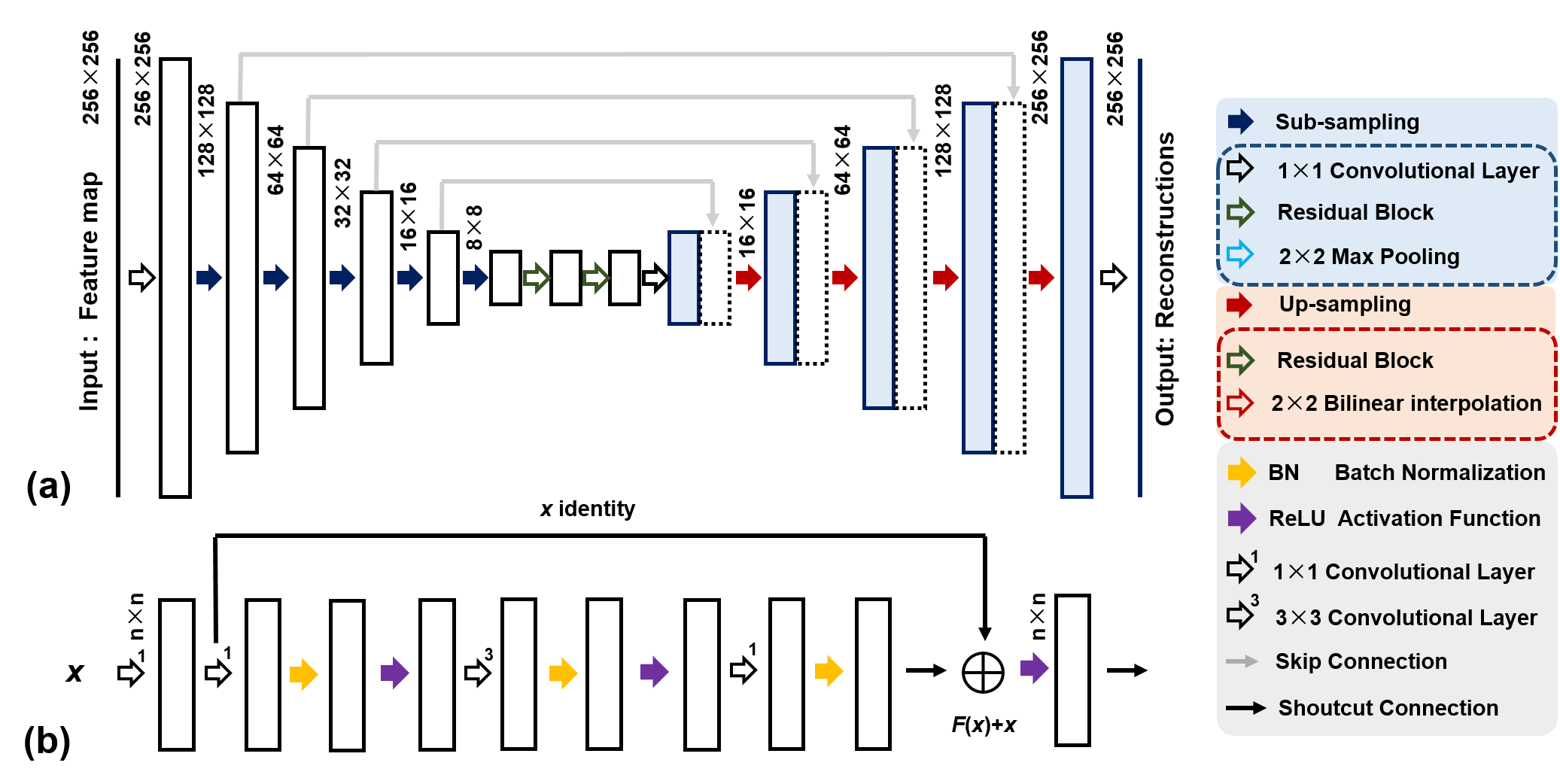}
\caption{The structure of U-Net backbone and the residual block module. (a) the structure of U-Net in which the block represents the processed tensors in each layer, (b) the structure of the residual block and its data format.}
\label{fig:unet}
\end{figure*}

In the last step of solving the inverse problem in NLOS reconstruction, we adopt a U-Net as the backbone structure, as shown in Figure \ref{fig:unet}. Skip connection is used to perform multi-scale feature fusion, and a residual block is added to avoid gradient explosion during training.

We design adaptive data loaders for three different paradigms of sensors data, as shown in Figure \ref{fig:EPNP}. Screenshots of the diffusion spot and event TS maps are loaded directly as 2d image form to the network, serving as the frame-based intensity data and event-based feature map. It’s worth noting that the fusion form is the combination of frame-based image and even-based TS map, which is loaded as a concatenated tensor.

For the pre-training session, we trained our model on the simulation training set of each data paradigm with an adaptive moment (Adam) estimation optimizer for 200 epochs, respectively. For the enhanced reconstruction session, we load the pre-trained model with the test set of each data format, and fine-tune the model for approximately 100 epochs before sending it through the decoder for reconstructions.

To conclude, the event-based sampling paradigm together with the EPNP framework provides a fast and effective gateway for passive NLOS imaging of moving objects or dynamic scenes, as a practical application of intelligent sensor.

\section{Experiments and Results}

As proof of concept, we demonstrate the performance of EPNP prototype through simulation and real-shot experiments.

\subsection{Experimental settings}

\begin{figure}[htbp]
\centering
\includegraphics[width=0.8\linewidth]{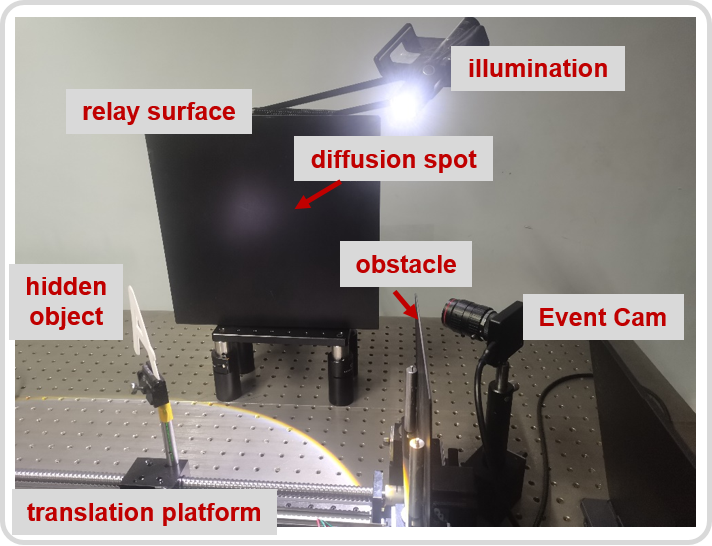}
\caption{The settings of real world experiment. An event camera (CeleX-MIPI) is used as the sensor for NLOS data acquisitions of moving objects on a translation platform.}
\label{fig:exp_setting}
\end{figure}

The experimental settings are shown in Figure \ref{fig:exp_setting}, an illuminated target ‘A’ is moving from left to right under indoor ambient light, controlled by a motorized translation platform and blocked by the obstacle.

The targets used in the real-world experiment are fabricated by laser printing acrylic material in the shape of alphabet ‘A-Z’ and number ‘0-9’ in Comic font, with a size of $ 8cm \times 8cm $, which translates at a speed of 2 cm per second. A frosted aluminum fender is used as the relay surface, which provides homogeneous scattering, and the moving target is placed about 40 cm from the fender. We use a torch to illuminate the hidden target and use an event camera (CeleX V-MIPI) to record the moving diffusion spot on the relay surface. Event-based data and frame-based data are acquired with different modes of the CeleX camera. Specifically, we obtain the event stream data in event-intensity (EI) mode and get frame series at a frame rate of 100 frames per second in full-picture (F) mode, which recorded the dynamic diffusion spot corresponding to the movement of the hidden object.

\subsection{Simulation experiments and results}

To evaluate the performance and effectiveness of the proposed EPNP prototype, simulation experiments are conducted to pre-train the physical embedded model and verify the enhancement with event-based vision. 

\begin{figure*}[htbp]
\centering
\includegraphics[width=0.75\linewidth]{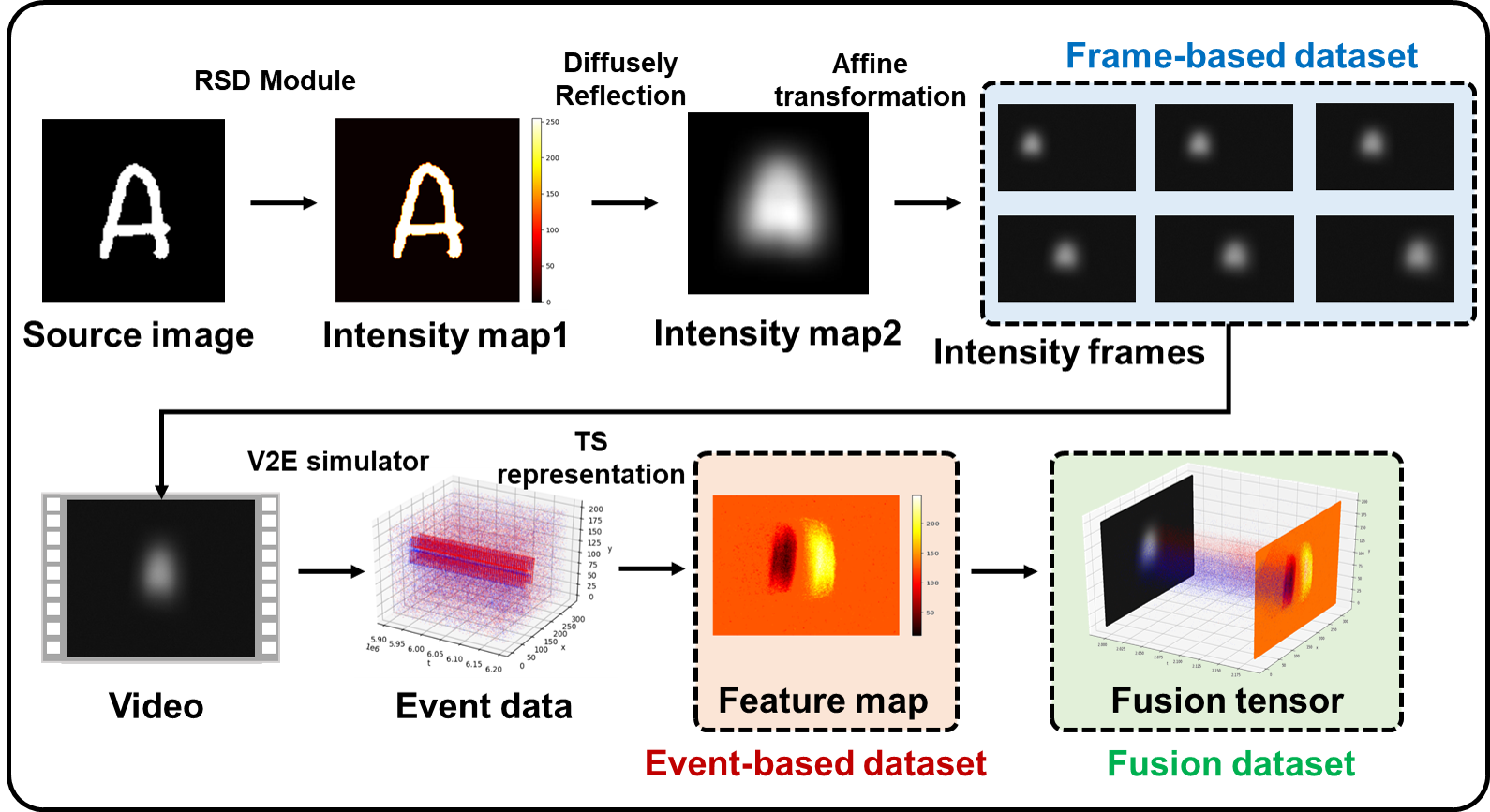}
\caption{The detailed pipline of simulation, datasets of three data paradigms are obtained consecutively.}
\label{fig:simu_process}
\end{figure*}

First, we perform simulations according to the forward model and corresponding experimental settings. The simulation pipeline of generating the light field on the relay surface is illustrated in Figure \ref{fig:simu_process}. 
Based on the source image including alphabet characters selected from the NIST dataset \cite{NIST:58} and hand-crafted comic digits, we generate the intensity distribution of diffusion spots on the relay surface at different positions along with the target movement. The simulation process for the fundamental frame-based dataset is conducted through RSD module, diffusely reflection simulation, and affine transformation.
Specifically, in RSD module, we set the diffraction propagation distance as 500 mm, the center wavelength as 500 nm and perform RSD simulation by convoluting with the transfer function in the form of angular spectrum. Similarly, in simulation of diffusely reflection, we adopt the equivalent tPSF calculated with parameter value $t=2, <l>=0.8$, and generate the intensity distribution after interacting with the relay surface by convolution. Affine transformation operations are performed to consider the geometric relationships between the sensor and its FoV. Moreover, random noise and gain factors are also included in the simulation.

After obtaining the fundamental frame-based dataset, the event-based and fusion datasets are established consecutively. We use the V2E simulator to generate event data from the videos synthesized by simulated intensity frames, perform TS calculation for dynamic feature extraction, and obtain the TS feature map as event-based dataset. For fusion dataset, we concatenate the TS map with intensity frame corresponding to the concerned event data, and save it as a feature tensor. The generated simulation datasets are presented in Table \ref{tab:simu_dataset},

\begin{table}[htbp]
\centering
\caption{\bf Names and data volume of the simulated datasets.}
\begin{tabular}{cc}
\hline
train \& validation & test \\
\hline

\toprule
    \multirow{3}{*}{NIST letters \quad $ 39 \times 26 (40)$} & \multicolumn{1}{c} {NIST letters \quad $ 1 \times 26 (40)$}  \\
    \cline{2-2}
    & Comic letters \quad $ 1 \times 26 (40)$ \\
    \cline{2-2}
    & Comic nums \quad $ 1 \times 10 (40)$ \\
\hline
\end{tabular}
  \label{tab:simu_dataset}
\end{table}

\noindent where the training set contains 1014 different types of alphabet character in the NIST dataset \cite{NIST:58}, and the testing set include 26 letters each from the NIST and Comic dataset. The number ‘40’ in brackets indicates that we generate data at 40 equidistant positions through target movement.

\begin{figure*}[htbp]
\centering
\includegraphics[width=0.95\linewidth]{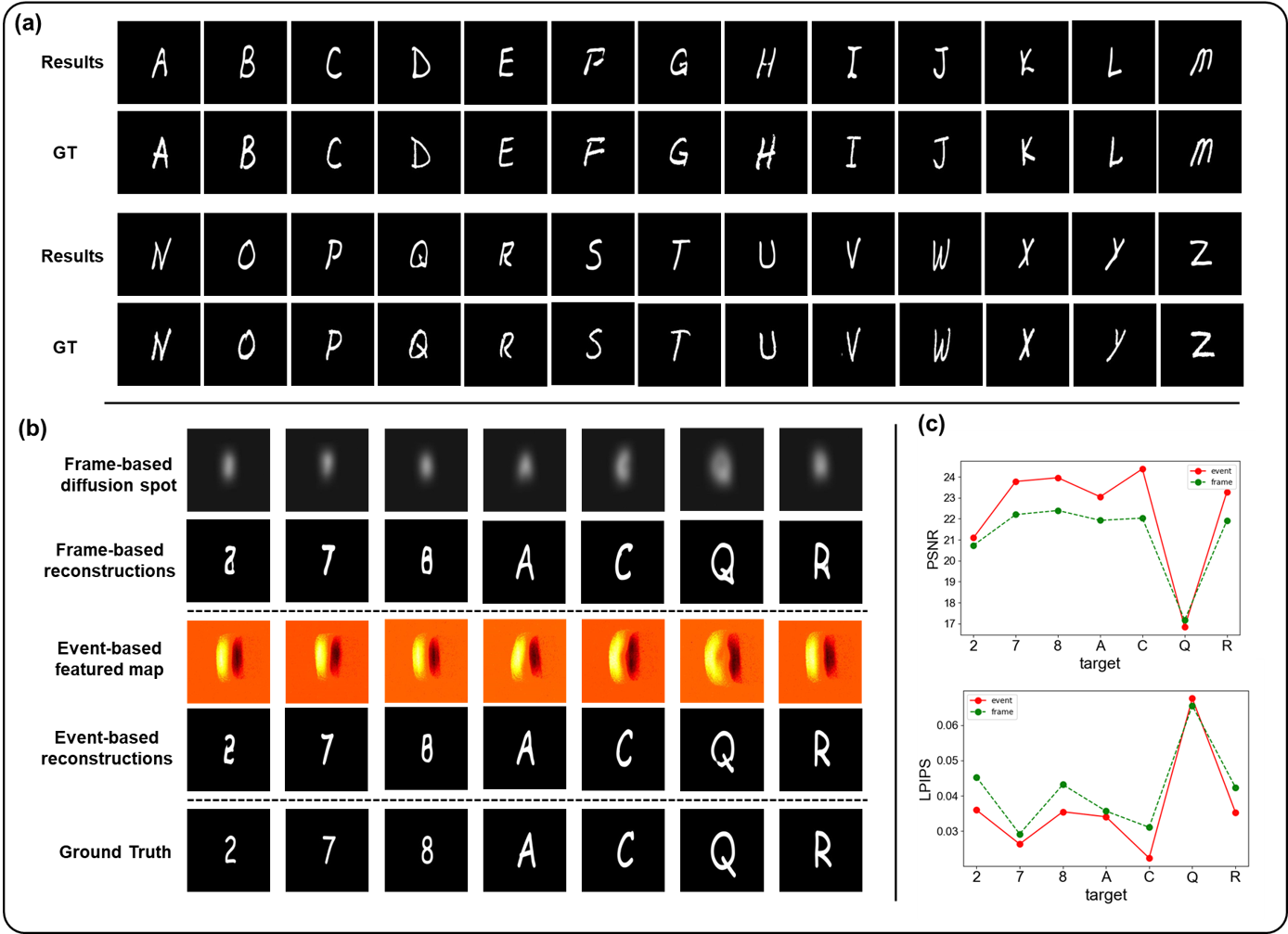}
\caption{The Simulation experimental results through EPNP prototype, (a) test results on NIST test set, (b) comparisons of event-based and frame-based reconstructions on Comic test set, (c) plots of evaluation metrics.}
\label{fig:simu_result}
\end{figure*}

Finally, with the prepared simulation data sets, we conduct simulation experiments by comparing the event-based and frame-based reconstructions, as shown in Figure \Ref{fig:simu_result}.

We train the simulated training set with the designed UNet structure in Figure \Ref{fig:unet}, utilizing an adaptive moment (Adam) estimation optimizer implemented on PyTorch 1.7.0 with an Nvidia RTX 3090 GPU for 200 epochs. A learning rate of $10^{-4}$ is used for the first 100 epochs, whereas it varies to $5 \times 10^{-5}$ for the remaining epochs, with batch size of 16. The loss function $l$ is crafted as a combination of Mean Absolute Error (MAE) and Structure Similarity Index Measure (SSIM):

\begin{equation}
l = L_1 ( \hat{x}, x_0 ) + 2 \cdot \frac{1}{SSIM( \hat{x}, x_0)} ,
\label{eq:loss}
\end{equation}

\noindent where $\hat{x}$, $x_0$ denotes the reconstruction and ground truth, $L_1$ represents L1 loss. The weight factor of the SSIM loss is selected as 2 to better balance these two components when training.

Both the event-based and frame-based datasets are pre-trained, and the test results on the event-based simulated testing set (NIST) are shown in Figure \ref{fig:simu_result} (a), which verifies the generalization of the pre-trained network. The simulated testing set (Comic) is considered as another data distribution and is used to examine the transfer ability of physical embedding enhancement induced by the pre-trained model on a sufficient simulation dataset. We load the pre-trained model and the testing set (Comic) of frame-based and event-based formats into the framework, respectively, and fine-tune the model for 100 epochs with learning rate of $10^{-5}$ and batch size of 16, implemented on same configurations. The test results are shown in Figure \ref{fig:simu_result} (b), and the evaluation metrics of event-based and frame-based methods are compared in Figure \ref{fig:simu_result} (c). We use peak signal to noise ratio (PSNR) and learned perceptual image patch similarity (LPIPS) \cite{Zhang:59} with a VGG core to assess the reconstruction quality. PSNR quantifies pixel-wise distortion, and LPIPS extracts features through neural networks and calculates feature differences for effective evaluation, which is more in line with human perception. It’s worth noting that a lower LPIPS indicates better reconstructions, which is different from traditional evaluation indices such as PSNR.

It can be seen from the test results that event-based method performs better than its frame-based counterpart when using the EPNP prototype to fine-tune and transfer between different data distributions, especially the reconstructions of Comic numbers.

\subsection{Real-shot experiments and results}

For the experimental proof of the EPNP prototype, we construct the experimental setup, as shown in Figure \ref{fig:exp_setting}. Real world experiments are conducted to validate the superiority of the EPNP prototype compared with E2E data-driven method. 

We use an event camera (CeleX-V MIPI) to record the moving diffusion spot on the relay surface in event-based mode (EI) and frame-based mode (F) respectively. The collected data under the EI and F modes are converted into real-shot datasets after calibration to the ground truth by the initial time stamp. The targets we used are flat acrylic characters of alphabet letters and numbers in Comic font. The distance between the target and relay surface are set as 30 cm (near) and 50 cm (far) for two sets of experiments, to validate the robustness. 

To the best of our knowledge, we establish the first Comic-NLOS dataset for event-based NLOS captures, which contains 2880 images in each of the three data paradigms, i.e. frame, event and fusion. The fabrication process of the real-shot datasets of each data paradigm is the same as that of their simulation counterparts. The naming and composition of the datasets are listed in Table \ref{tab:real_dataset},

\begin{table}[htbp]
\centering
\caption{\bf Names and data volume of the real-shot datasets.}
\label{tab:real_dataset}
\begin{tabular}{lccc}
\hline
Name&frame& event& fusion\\
\hline
Comic letters near&$1 \times 26(40)$& $1 \times 26(40)$& $1 \times 26(40)$\\
Comic nums near&$1 \times 10(40)$& $1 \times 10(40)$& $1 \times 10(40)$\\
Comic letters far&$1 \times 26(40)$& $1 \times 26(40)$& $1 \times 26(40)$\\
Comic nums far&$1 \times 10(40)$& $1 \times 10(40)$& $1 \times 10(40)$\\
\hline
\end{tabular}

\end{table}

\noindent where the 'Comic letters' dataset contains 26 alphabet characters, the 'Comic nums' dataset contains 10 number digits, and the suffix of ‘near’ and ‘far’ supplies the distance from the target to the relay surface. The captured data of each target are produced into 40 featured images, which represent the movement of the diffusion spot at different positions.  

\begin{figure*}[htbp]
\centering
\includegraphics[width=0.88\linewidth]{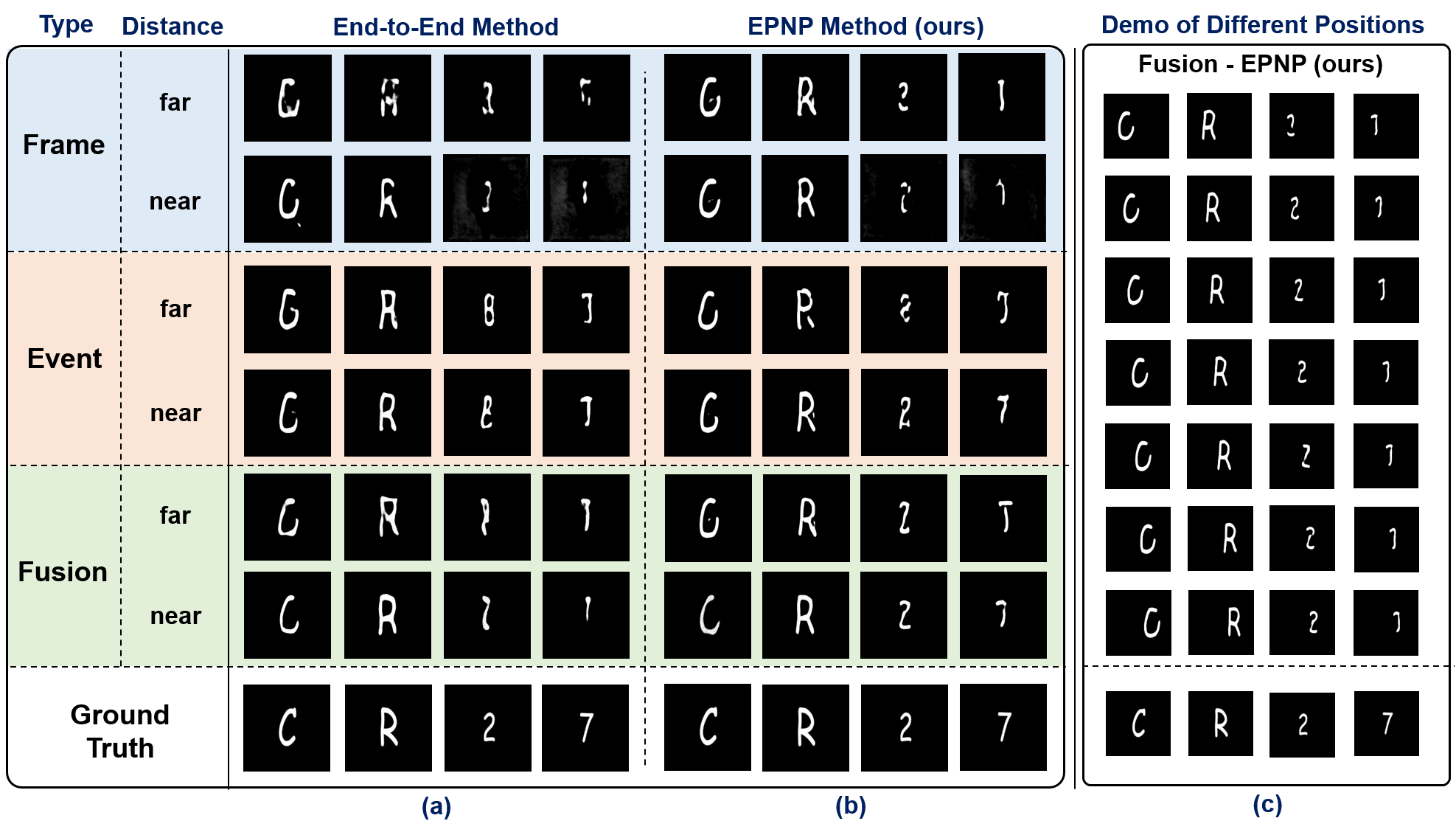}
\caption{The Real-shot experimental results, 'near' (30 cm) and 'far' (50 cm) refer to the distance between targets and relay surface, respectively. (a) test results through E2E approach, (b) test results through EPNP prototype, (c) a demonstration for reconstructions at different positions for fusion test set through EPNP prototype.}
\label{fig:real_result}
\end{figure*}

After adopting the pre-trained model on simulation dataset for physical embedding, we load the real-shot dataset for fine-tuning through the EPNP framework. The pre-trained model is trained on the simulated NIST dataset only, and we fine-tune the network with both Comic letters and Comic nums datasets of the three data paradigms.
The test results of real-shot experiment using E2E data-driven method and EPNP framework are shown in Figure \Ref{fig:real_result}(a)(b), respectively, and the reconstructions of target movement at different positions with fusion format data through EPNP are demostrated in Figure. \Ref{fig:real_result}(c).
We also use PSNR and LPIPS as the evaluation metrics, and the average scores of each target reconstruction are listed in Table \Ref{tab:real_results}.
The best and second-best indicators are marked in red and blue respectively. It is obvious that the event-based paradigm and fusion paradigm perform better than the frame-based counterpart, which verifies the superiority of event-based applications in passive NLOS imaging for moving objects.

\begin{table*}[htbp]
\centering
\caption{\bf PSNR and LPIPS results of the reconstruction of Comic test set in real-shot experiments.}
\label{tab:real_results}
\begin{tabular}{c|ccc|ccc|ccc|ccc}
\hline
 \multirow{3}{*} {Target}& \multicolumn{6}{c}{Near group}& \multicolumn{6}{c}{Far group}\\
\cline{2-13}
 & \multicolumn{3}{c}{PSNR/dB $(\uparrow)$} & \multicolumn{3}{c}{LPIPS $(\downarrow)$} & \multicolumn{3}{c}{PSNR/dB $(\uparrow)$}& \multicolumn{3}{c}{LPIPS $(\downarrow)$}\\
\cline{2-13}
   & frame& event&fusion  & frame& event&fusion & frame& event&fusion  & frame & event&fusion \\
\hline
 C& 15.88&  \textcolor{blue}{17.14}&  \textcolor{red}{17.66}  & 0.071& \textcolor{red}{0.055}&\textcolor{blue}{0.070} & 15.97& \textcolor{blue}{16.62}&\textcolor{red}{17.16}& 0.069& \textcolor{red}{0.066}&\textcolor{blue}{0.067} 
\\
 R& 16.93&  \textcolor{blue}{17.46}&  \textcolor{red}{19.18}& 0.066& \textcolor{red}{0.054}&\textcolor{blue}{0.056} & \textcolor{blue}{17.09}& 16.93&\textcolor{red}{18.88}& 0.072& \textcolor{blue}{0.061}&\textcolor{red}{0.060} 
\\
 2& 19.58&  \textcolor{blue}{20.64}&  \textcolor{red}{21.12}& 0.169& \textcolor{blue}{0.045}&\textcolor{red}{0.032} & 18.46& \textcolor{red}{19.88}&\textcolor{blue}{18.49}& 0.051& 0.051&\textcolor{red}{0.047} 
\\
 7& 19.34&  \textcolor{red}{20.47}&  \textcolor{blue}{19.69}& 0.221& \textcolor{blue}{0.053}& \textcolor{red}{0.051} & 17.29& \textcolor{red}{17.85}&\textcolor{blue}{17.44}& 0.071& \textcolor{red}{0.058}&\textcolor{blue}{0.065} \\
 \hline
\end{tabular}
\end{table*}

\begin{figure*}[htbp]
\centering
\includegraphics[width= 0.85\linewidth]{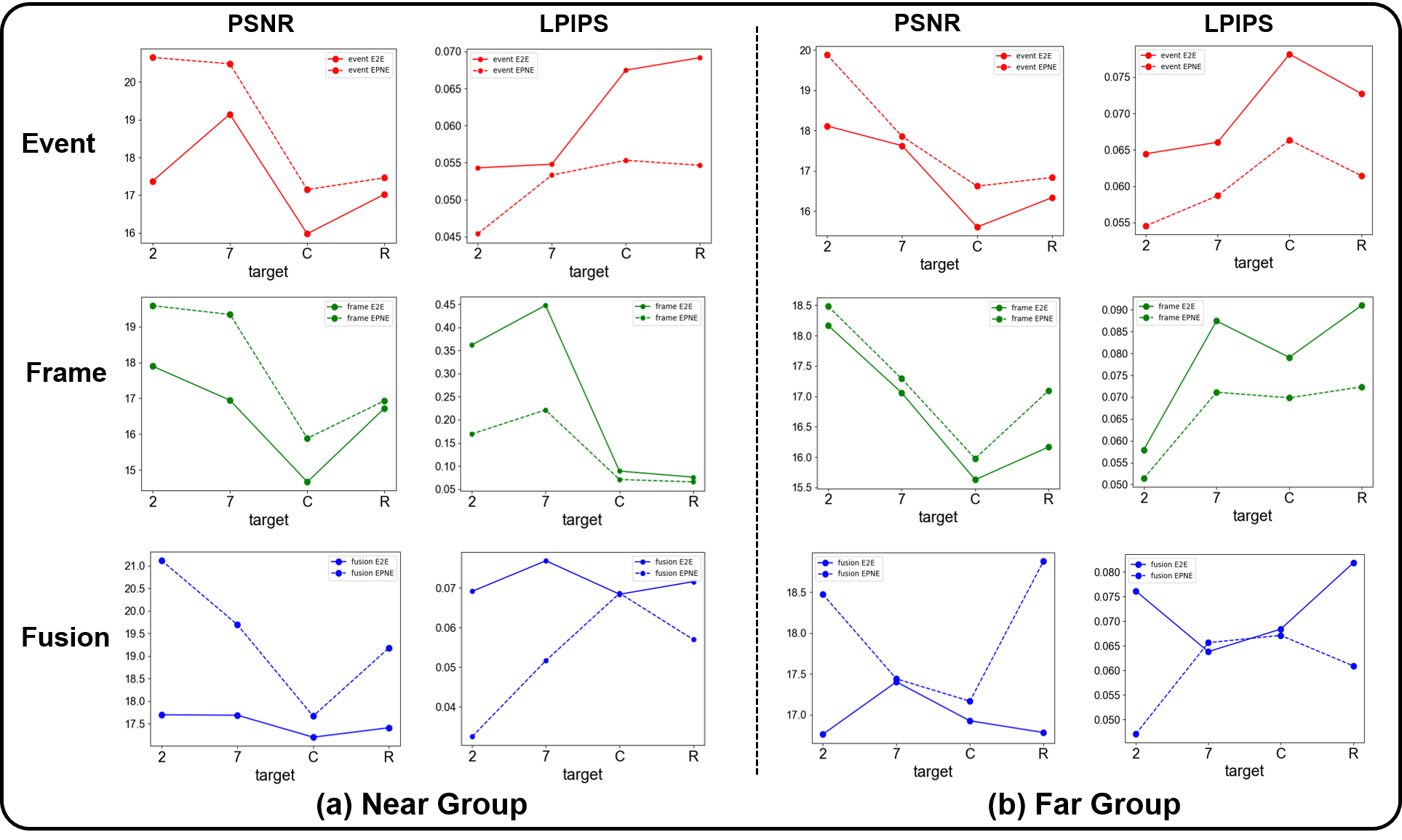}
\caption{The plots for evaluation metrics comparison of the proposed EPNP method with E2E data-driven method, where the full lines represent the E2E methods and the dotted lines represent the EPNP method.}
\label{fig:real_result_plot}
\end{figure*}

In addition, comparisons of reconstructions through E2E and EPNP method are shown in Figure \ref{fig:real_result_plot}, with three sensors data paradigms respectively. The reconstructions of the EPNP method perform better than that of the E2E method in most cases, which demonstrates the importance of embedding physical restrains into the data-driven approach in passive NLOS imaging.

\section{Discussions}

\subsection{Results analysis}

From the comparison of objective evaluation indicators between the three data paradigms listed in Table \ref{tab:real_results}, the event and fusion paradigm outperforms the frame paradigm on both of the alphabet and the number targets in each distance group. A better performance on PSNR indicates more accurate restoration of textures, and the superiority on LPIPS expresses more conformity with human eye perception, which validate the enhancement brought by event-based sensing for NLOS moving target reconstruction. 

Deeply exploring the relationship between information representation and reconstruction results, we find the unique advantage of each type of sensors data paradigm in its applications. It can be seen from the real-shot experiment results that through the EPNP framework, event-based information performs well in reconstructing the texture details, whereas frame-based information is more effective in reconstructing the morphological basis of NLOS moving targets. As a result, the fusion of event-based feature with traditional frame-based intensity information is an effective approach to expand and enhance the applicability in practical scenarios \cite{Cao:60}.

The curves of PSNR and LPIPS with three dataset paradigms using E2E and EPNP method are compared in Figure. \ref{fig:real_result_plot}. The proposed EPNP method represented by the dotted line achieves better indices on both metrics than the E2E method represented by full lines. The margin of evaluation scores between alphabet targets and number targets is mainly caused by the difference in data distribution and font size, since the targets of 'Comic nums' are smaller than that of 'Comic letters'. 
On the other hand, the correctness of reconstructions on Comic nums ‘2’ and ‘7’ through EPNP compared with the over-fitting results through E2E method demonstrate the generalization ability of EPNP when the real-shot dataset is relatively small. Taking the EPNP prototype, more physical constraints from the simulation datasets are embedded and the reconstruction is enhanced in real-world experiments. Also, the implementation of event-based sensor enhances the reconstruction performance in NLOS dynamic scenes.

\subsection{Contributions}

In this study, we constructed a passive NLOS system with illuminated moving targets, where event-based and frame-based data are captured for real-shot dataset fabrication.
A plug and play prototype EPNP is proposed, leveraging the enhancement provided by event-based vision and physical constrained pipeline. The established scheme acts as a practical application of novel sensors in computational imaging.

Since the main challenge of passive NLOS imaging for moving objects is optimizing the inverse problem under the strong spatio-temporal aliasing interference in traditional sensor acquisition, the proposed event-enhanced framework can mitigate blindness with the support from physical basis and in-sensor computing. In addition, the involvement and fusion of event-based vision provides more powerful dynamic feature extraction and representation capabilities, which enhances the reconstruction performance for moving objects in passive NLOS imaging.

\subsection{Limitations and Prospects in application}

Event paradigm detection shows superior dynamic feature extraction and in-sensor computing capabilities, however, the absolute intensity information is lost during the data acquisition process. Although TS representation can use spatio-temporal correlation features to reconstruct targets, offloading the importance of absolute intensity information to some extent, without the support of frame-based information, many morphological features are neglected. This leads to shortcomings when reconstructing solid targets or full scenery, in which only contours and textures are reconstructed. Therefore, the application of event-based vision through the physical embedded passive NLOS prototype EPNP still has limitations in static and blocky object imaging, and we could not rely solely on event paradigm information. To solve this, the fusion of these two information dimensions are required to ensure better performance and more flexibility in generalization.  

We believe that the application of event-based vision in computational imaging opens the door for device-dominated in-sensor computing \cite{Zhou:61} and sensing methods. The proposed EPNP prototype will guide the optimization of solving the inverse problem by simulated physical embedding for enhancement in restorations. In future work, more complicated targets of various data distributions can be added to the simulation part with a more accurate forward model expression to further enhance the performance and generalization. In addition, the structure of the embedding module and feature representation format of different information dimension can be studied to ensure high quality pre-training. The application of event-based vision together with the EPNP prototype will not limited to passive NLOS imaging for moving objects. It can be promoted in more applications where dynamic feature extraction is required or under complex and harsh ambient light field environments. Besides the event feature embedding, more dimensions of the light field could be simulated and captured by photonic sensors through the EPNP prototype with more suitable network structures.

\section{Conclusions}

Traditional frame-based data acquisition paradigm exhibits shortcomings in passive NLOS moving target imaging, and the E2E data-driven reconstruction method limits the generalization and applicability with a limited dataset. The movement of hidden targets further increase the blindness of the ill-posed inverse problem in temporal dimension. This paper puts forward a step towards overcoming these limitations. Leveraging the sampling paradigm of dynamic vision sensors, an event-based enhanced passive NLOS imaging method with simulated physical embedding is proposed. The EPNP prototype promises to enable high-quality reconstruction of moving targets occluded by obstacles and ensure certain generalization among data distributions. 

Specifically, the principles and procedures of the simulation dataset fabrication and target reconstruction with the EPNP framework are illustrated in detail, and the effectiveness is visualized and quantitatively evaluated for comparison with the E2E data-driven framework. The superiority of the proposed method in terms of performance is validated through simulation and real-world experiments. We also discuss the limitations and possible application prospects for future work, which provides a new gateway for physical embedded NLOS reconstruction with extra enhancement with in-sensor computing and feature extraction of different information paradigms. 

When artificial intelligence (AI) meets photonics, the effectiveness of information acquisition using novel photonic sensors, coupled with the rationality of information utilization through AI is essential for enhancing the solving process of ill-posed problems. The proposed plug and play structure EPNP has great potential for facilitating further research on physically enhanced and sensor-dominated dynamic scene computational imaging applications beyond passive NLOS imaging.

\appendices


\section*{Acknowledgment}

The authors thank Dr. Yutong He and Dr. Hao Shi for insightful discussions in event-based computational imaging.

\ifCLASSOPTIONcaptionsoff
  \newpage
\fi

\end{document}